\documentclass[12pt]{article}
\pdfoutput=1

\usepackage{jheppub}
\usepackage{amsmath,amssymb,euscript,array,mathrsfs}
\usepackage{slashed}

\usepackage{epsfig}

\def\a{\alpha}
\def\b{\beta}
\def\c{\gamma}
\def\d{\delta}

\def\l{\lambda}
\def\m{\mu}
\def\n{\nu}

\def\s{\sigma}

\def\w{\omega}

\def\D{\Delta}

\def\tr{{\rm tr}}

\def\Dbarslash{\,\,{\raise.15ex\hbox{/}\mkern-12mu {\bar D}}}
\def\Dslash{\,\,{\raise.15ex\hbox{/}\mkern-12mu D}}
\def\delslash{\,\,{\raise.15ex\hbox{/}\mkern-9mu \partial}}
\def\delbarslash{\,\,{\raise.15ex\hbox{/}\mkern-9mu {\bar\partial}}}

\def\half{\frac{1}{2}}
\def\thalf{\tfrac{1}{2}}

\def\rta{\rightarrow}
\def\tr{{\rm tr}}

\def\Im{{\rm Im}}
\def\Re{{\rm Re}}

\def\MOL{{\textsf{MOL }}}
\def\EXP{{\textsf{EXP }}}
\def\HEL{{\textsf{HEL }}}
\def\LAB{{\textsf{LAB }}}
\def\SUN{{\textsf{SUN }}}
\def\TT{{\cal T}}
\def\VV{{\cal V}}
\def\NR{{\rm NR}}
\newcommand{\wh}{\widehat}

\newcommand{\Hmol}{$\textrm{H}_2^{\,+}$}
\newcommand{\Hbarmol}{$\overline{\rm H}_2^{\,-}$}
\newcommand{\bs}{\boldsymbol}

\newcommand{\beq}{\begin{equation}}
\newcommand{\eeq}{\end{equation}}

\newcommand{\la}{\langle}
\newcommand{\ra}{\rangle}

\title{\begin{center}{\Large  Lorentz and \textsf{CPT} violation and the
hydrogen and antihydrogen molecular ions III -- rovibrational spectrum and the non-minimal SME}
\end{center}\vskip0.5cm}

\author{\vskip1cm
\large{Graham M.~Shore} }

\emailAdd{g.m.shore@swansea.ac.uk  }

\affiliation{\vskip0.8cm 
Centre for Quantum Fields and Gravity, Department of Physics, Swansea University, \\
Singleton Park, Swansea, SA2 8PP, UK
\vskip0.2cm
}

\date{\today}

\abstract{Rovibrational transitions in the hydrogen and  antihydrogen molecular ions
\Hmol and \Hbarmol offer the possibility of testing Lorentz and \textsf{CPT} symmetry to extremely
high precision, in principle attaining $O(10^{-17})$.
In this paper, the third in a series, we give a comprehensive derivation of the rovibrational spectrum 
of \Hmol and \Hbarmol in the SME, an effective quantum field theory incorporating Lorentz and
\textsf{CPT} violation. New developments described here include a complete analysis of the 
molecular dynamics from first principles in terms of the spherical tensor representation of the SME couplings, 
the systematic extension of our previous results to the non-minimal SME, a full description of the 
quantum number dependence of the rovibrational energy levels in the spherical tensor formalism 
with both high and low background magnetic fields, and an extended discussion of sidereal and 
annual variations of transition frequencies arising from both rotations and Lorentz boosts.
The resulting sensitivity of the rovibrational spectrum to an extended range of SME couplings,
together with the ability to isolate their individual effects using the quantum number 
dependence of the transition frequencies, enhances the opportunities to detect Lorentz and 
\textsf{CPT} symmetry breaking through rovibrational spectroscopy of \Hmol and \Hbarmol.
}

\notoc

\begin{document}

\maketitle

\setlength{\parskip}{10pt}

\newpage

\section{Introduction}\label{sect 1}

Lorentz invariance and \textsf{CPT} symmetry are fundamental elements of the local relativistic 
quantum field theories that provide our modern understanding of elementary particle physics. 
It is therefore important to find ways to test them experimentally at the highest levels of precision 
\cite{Charlton:2020kie}.
Rovibrational spectroscopy of the hydrogen and antihydrogen molecular ions \Hmol and \Hbarmol\,
is an especially promising direction, offering the possibility of direct tests of Lorentz and 
\textsf{CPT} symmetry at precisions potentially of $O(10^{-17})$ \cite{SchillerCP}. 

This paper is the third in a series \cite{Shore:2024ked,Shore:2025zor} (referred to from now on as 
Papers 1 and 2) in which we study the rovibrational spectrum of \Hmol and \Hbarmol\,
in an effective theory of Lorentz and \textsf{CPT} violation, the standard model extension (SME) 
\cite{Colladay:1998fq}.
New developments here include a full derivation of the molecular dynamics from first principles in terms
of the spherical tensor representation of the SME couplings, the systematic extension of our previous results
to the non-minimal SME \cite{Kostelecky:2013rta}, 
a full description of the quantum number dependence of the rovibrational 
energy levels in the spherical tensor formalism at both high and low background magnetic fields, 
and an extended discussion of sidereal and annual variations of transition frequencies arising
from both rotations and Lorentz boosts.

Significant advances in experimental techniques have allowed rapid progress in the rovibrational spectroscopy
of \Hmol in RF traps, with current measurements already achieving $O(10^{-12})$ precision \cite{SAS2024}.
In parallel, recent developments towards transporting antiprotons away from the CERN AD/ELENA
facility to a dedicated off-site spectroscopy laboratory, together with 
progress in understanding routes to synthesise \Hbarmol\, \cite{Zammit, Myers, Zammit:2025nky}
and the increasing precision achieved by the ALPHA collaboration in $1S$\,-\,$2S$ spectroscopy 
of atomic antihydrogen \cite{Ahmadi:2018eca, Baker:2025ehs}, 
are opening a path towards future rovibrational spectroscopy of \Hbarmol.
This would enable direct high-precision tests of \textsf{CPT} by comparison of the spectra of
\Hmol\, and \Hbarmol\, in a common experiment.

The SME Lagrangian \cite{Colladay:1998fq} is parametrised by a set of Lorentz tensor couplings to operators formed 
from the standard model fields, in contrast to the scalar couplings of conventional Lorentz-invariant QFTs,
{\it viz.}
\begin{align}
{\cal L}_{\rm SME} \,=\,&\frac{1}{2} \sum_w \int d^4 x~\Bigl[ 
\bar\psi\left(i \c^\m \partial_\m \,-\, m\right)\psi
\,-\, a^{\,w}_\m \,\bar\psi \c^\m \psi \,+\, i c^{\,w}_{\m\n}\,\bar\psi \c^\m \partial^\n\psi \,+\,
a^{\,w}_{\m\n\l}\, \bar\psi \c^\m \partial^\n \partial^\l\psi     \nonumber \\[3pt]
& -\, b^{\,w}_\m\, \bar\psi\c^5 \c^\m \psi 
\,+\, i d^{\,w}_{\m\n}\, \bar\psi \c^5\c^\m \partial^\n \psi  
\,-\,  \tfrac{1}{2} H^{\,w}_{\m\n}\, \bar\psi\s^{\m\n}\psi  \,+\, 
\tfrac{1}{2}i g^{\,w}_{\m\n\l}\, \bar\psi\s^{\m\n} \partial^\l \psi
~+~\ldots~ \Bigr] \nonumber \\[3pt]
&+~~ {\rm h.c.}
\label{a1}
\end{align}
where $w$ denotes the particle type, so here the sum is over the electron and the two protons
in the molecular ion.
The dots represent further couplings to higher-dimensional, non-renormalisable operators,
of which $a^{\,w}_{\m\n\l}$ is the first example. 
The $a, \,b$ and $g$ couplings also violate \textsf{CPT} symmetry, while $c, \,d$ and $H$
are Lorentz violating but preserve \textsf{CPT}.  It is often useful to think of these couplings
as VEVs of vector and tensor fields, perhaps of some more fundamental theory.

To analyse atomic and molecular systems, we need the non-relativistic Hamiltonian derived 
from this, and in this paper we use from the outset the very convenient formalism (see \cite{Kostelecky:2013rta}
for an extensive description) in which the couplings are represented as spherical tensors. 
This is especially well-suited to the symmetries of atoms and molecules and makes the extension
to the non-minimal SME, including non-renormalisable operators, particularly natural.
This Hamiltonian, for para-\Hmol, may be written as
\begin{align}
H_{\rm SME} ~=~ &- \sum_{njm} \,p_e^n \,Y_{jm}(\hat{\bs{p}_e})\,\VV_{njm}^{e\,\NR} \,-\,
\sum_{njm} \,p_1^n \,Y_{jm}(\hat{\bs{p}_1})\,\VV_{njm}^{p\,\NR} \,-\,
\sum_{njm} \,p_2^n \,Y_{jm}(\hat{\bs{p}_2})\,\VV_{njm}^{p\,\NR} \nonumber \\[6pt]
&- \s_{\HEL}^r \,\sum_{njm}\,p_e^n \,Y_{jm}(\hat{\bs p}_e) \,\TT_{njm}^{e\,\NR\,(0B)}  \nonumber \\[6pt]
&+ \s_{\HEL}^{\pm} \,\sum_{njm}\, p_e^n\,\,{}_{\pm 1}Y_{jm}(\hat{\bs p})\,\big(
\pm \TT_{njm}^{e\,\NR\,(1B)} \,+\,i \TT_{njm}^{e\,\NR\,(1E)}\big) \ ,
\label{b11}
\end{align}
where $\s_{\HEL}^{r,\pm}$ denote the components of the electron spin in the helicity frame (\textsf{HEL})
associated with its momentum. Here,\,\,${}_sY_{jm}$ are the spin-weighted spherical harmonics, defined in
section \ref{sect 3}. The dictionary relating the spherical and Lorentz tensor couplings is given
in \cite{Kostelecky:2013rta} and Paper 2, with further identities shown here where necessary.

The breaking of Lorentz invariance means that frames of reference play an important r\^ole in our analysis.
First of these is the reference frame (\textsf{MOL}) aligned with the molecular axis. This is used to analyse 
the electron dynamics, which in the standard Born-Oppenheimer framework determines the inter-nucleon
potential. A notable point here is that even in the electron $1s\s_g$ ground state, the molecule only has
cylindrical and not full spherical symmetry, which results in sensitivity to the electron spherical tensor 
couplings $\VV_{njm}^{\,e}$ and $\TT_{njm}^{\,e}$ with $j\neq 0$.  Our results for the rovibrational energies 
themselves are then calculated in the \textsf{EXP} frame, with quantisation axis aligned with the background
magnetic field.
An important r\^ole is also played by a solar-centred frame (\textsf{SUN}), which has become the 
standard in which to compare constraints on the SME couplings arising from different experiments 
\cite{Kostelecky:2008ts}.
This frame is also the basis for our discussion of transition frequency variations from rotations
and Lorentz boosts.

In Papers 1 and 2 we have given an extensive description in the Born-Oppenheimer framework
of how the rovibrational energies are determined from the SME-modified inter-nucleon potential 
$V_M(R) \,+\,V_{\rm SME}^e(R,\boldsymbol{S})$, described in section \ref{sect 2},
together with the direct nucleon (proton) contribution $\D E_{\rm SME}^n$. 
The rovibrational energy spectrum is expressed as an expansion in powers
of $(v + \tfrac{1}{2})$ and $N(N+1)$ as follows:
\begin{align}
E_{v NM_J} ~=~ &V_{\rm SME}^e \,+\,(1 + \d_{\rm SME}^e + \d_{\rm SME}^n)\,(v+\thalf) \,\w_0  \,
-\,  (x_0 + x_{\rm SME}^e + x_{\rm SME}^n)\,(v+\thalf)^2 \,\w_0 \nonumber \\[5pt]
&+\, (B_0 + B_{\rm SME}^e + B_{\rm SME}^n) \,N(N+1) \,\w_0   \,  \nonumber \\[5pt]
&-\,  (\a_0 +\a_{\rm SME}^e + \a_{\rm SME}^n)\,(v + \thalf) N(N+1)\,\omega_0  \, \,+\,\ldots
\label{aE}
\end{align}
where $\w_0$ is the fundamental vibration frequency.  
Each coefficient of $\w_0$ is an expansion
in powers of a small parameter $\l$, parametrically of $O(\sqrt{m_e/m_p})$, with the leading terms 
in the coefficients $x_0, B_0, \a_0$ being of order $\l,\l, \l^2$ respectively.
In the rest of this paper, we calculate these coefficients in terms of the spherical tensor couplings
as perturbations on the hyperfine-Zeeman energy levels for para-\Hmol, determining their dependence
on the rovibrational and spin quantum numbers and their magnetic field dependence for both high and low
fields. This means our results may be applied directly both to current measurements with
\Hmol in low-field RF traps \cite{SAS2024} and future experiments with \Hbarmol, and \Hmol, 
confined in Penning traps \cite{SchillerCERN2026}.

A new feature in this work is the extended discussion of daily (sidereal) variations due to the Earth's
rotation, for which the the spherical tensor formalism is especially well-suited, and sidereal and 
annual variations due to Lorentz boosts with both the Earth's rotational and orbital velocities.
Here, the spherical tensor formalism is not in itself sufficient and we have to calculate directly
with the Lorentz tensor couplings, but here we exhibit the dictionary to translate these variations
back in terms of the spherical tensor couplings. This reveals a sensitivity to further SME couplings
with different values of both $n$ and $j$ quantum numbers, emphasising the need to exploit
these variations and not rely purely on time-averaged measurements.

One of the initial objectives of this paper was to bring together the approach of Papers 1 and 2
with the work \cite{Vargas:2025efi} (see also \cite{Vargas:2026chs}. This reference addresses the same topics, 
though in less generality and keeping only some of the leading terms in the rovibrational energies.
This means that the sensitivity to several SME couplings is not present in the results quoted there.
On the other hand, \cite{Vargas:2025efi} provides detailed explicit formulae for the frequency 
and rotational variations of a specific experimentally important rovibrational transition in \Hmol,
which we have checked are in agreement with the general results given here. 
Other earlier work on rovibrational spectroscopy and the SME includes \cite{Kostelecky:2015nma}
and \cite{Muller:2004tc}, the latter having particularly influenced Paper 1.

The paper is organised as follows. Section \ref{sect 2} gives a brief review of the modifications to the
Born-Oppenheimer approximation applied to the SME. Our main analysis of the inter-nucleon 
potential and rovibrational energy levels is given in sections \ref{sect 3.1}, \ref{sect 3.2} and \ref{sect 4},
for the electron spin-independent and spin-dependent couplings and the proton couplings respectively.
These results are quoted in the \textsf{EXP} frame, and translated to the \textsf{SUN} frame
in section \ref{sect 5},  where transition frequency variations due to rotations and Lorentz boosts 
are described.  Section \ref{sect 6} presents a brief summary and outlook on future work, 
while three appendices provide further technical details of our calculations.

\vskip1.5cm

\section{Born-Oppenheimer analysis in the SME}\label{sect 2}

We begin with a brief recap of the Born-Oppenheimer analysis incorporating Lorentz and
\textsf{CPT} violation described in Papers 1 and 2. 
In this approximation, the full Schr\"odinger equation is factorised into an ``electron
Schr\"odinger equation'', which determines the inter-nucleon potential $V_M(R)$, and
a ``nucleon Schr\"odinger equation'' from which we find the rovibrational energy eigenstates.
We restrict here to the experimentally relevant para-\Hmol, where the total nucleon spin is 0 
and the rotational angular momentum quantum number $N$ is even. 
As always the electron is taken in the $1s\s_g$ ground state.

Factorising the full wavefunction into electron and nucleon parts as\footnote{For kinematics and
notation, see Paper 1 and section 4 here..}
\begin{equation}
\Psi(\boldsymbol{R},\boldsymbol{r};M_s) ~=~ \Phi(\boldsymbol{R})\,\psi(\boldsymbol{r}; R)\, |M_s\rangle \ ,
\label{b1}
\end{equation}
the electron Schr\"odinger equation including the SME Hamiltonian is 
\begin{equation}
\bigg[- \frac{1}{2\hat{\mu}} \nabla_{\boldsymbol{r}}^2 \,+\, V_{mol}(R,r_{1e},r_{2e}) 
 ~+~ \tilde{H}_{\rm SME}^e(\bs{p};\bs{S})  
\,\bigg]\,\psi(\boldsymbol{r};R) \,|M_s\rangle  \,=\, 
E_e(\boldsymbol{R},\boldsymbol{S}) \,\psi(\boldsymbol{r};R)\, |M_s\rangle \ ,
\label{b2}
\end{equation}
where we understand $\boldsymbol{p} \rightarrow -i \nabla_{\boldsymbol{r}}$, and
$V_{mol}(R,r_{1e},r_{2e})$ is the electrostatic potential binding the molecule.
Here, the electron SME Hamiltonian $\tilde{H}_{\rm SME}^e(\bs{p};\bs{S})$
also contains a contribution from the spin-independent proton SME
couplings, as explained in Papers 1 and 2 and here in section \ref{sect 4}. 
This effect is included in the full inter-nucleon potential by the substitution given in (\ref{d10}).

The eigenvalues define the inter-nucleon potential:
\begin{equation}
E_e(\boldsymbol{R},\boldsymbol{S})~=~ V_M(R) ~+~ V_{\rm SME}^e(\boldsymbol{R},\boldsymbol{S}) \ ,
\label{b3}
\end{equation}
where 
\begin{equation}
V_{\rm SME}^e(\boldsymbol{R},\boldsymbol{S}) ~=~ \int d^3\boldsymbol{r} \, 
\psi^*(\boldsymbol{r};R) \, \tilde{H}_{\rm SME}(\bs{p};\bs{S})\, \psi(\bs{r};R) \ ,
\label{b4}
\end{equation}
is the expectation value of $\tilde{H}_{\rm SME}(\bs{p};\bs{S})$ in the 
electron $1s\s_g$ state. It depends on the expectation values $\langle p_a\, p_b\rangle$
of the electron evaluated in the \textsf{MOL} frame.

Inserting into the nucleon Schr\"odinger equation this gives
\begin{align}
&\bigg[-\frac{1}{2\mu} \nabla_{\boldsymbol{R}}^2 \,+\, V_M(R) 
\,+\, V_{\textrm{SME}}^e(\boldsymbol{R};\boldsymbol{S}) \,+\, H_{\rm SME}^p(\boldsymbol{P}) \bigg]
\Phi_{vNM_N}(\boldsymbol{R})\,|M_S\rangle 
\nonumber \\
&~~~~~~~~~~~~~~~~~~~~~~~~~~~~~~~~~~~~~~~~~~~~~~~~~~~~~~~~~~~~\,=\, 
E_{vNM_N M_S}\, \,\Phi_{vNM_N}(\boldsymbol{R})\,|M_S\rangle  \ ,
\label{b5}
\end{align}
with $\boldsymbol{P} = -i \nabla_{\boldsymbol{R}}$.  Here,
\begin{equation}
E_{vNM_N M_S} ~=~ \tilde{E}_{vNM_N M_S}\,+\,  \D E_{\rm SME}^n \ ,
\label{b6}
\end{equation}
with $\tilde{E}_{vNM_N M_S}$ found by setting $\Phi_{vNM_N}(\boldsymbol{R}) 
\,=\,\frac{1}{R} \phi_{vNM_N}(R) Y_{NM_N}(\theta,\phi)$ and solving the 1-dim 
rovibrational equation
\begin{align}
&\bigg[- \frac{1}{2\mu}\, \frac{d^2}{dR^2} \,+\, \frac{1}{2\mu R^2} N(N+1) \,+\, V_M(R) \,+\,
V_{\textrm{SME}}^e(R;\boldsymbol{S}) \bigg]\,\phi_{vNM_N}(R)\,  |M_S\rangle
\nonumber \\[10pt]
&=~ \tilde{E}_{vNM_N M_S}\,\, \phi_{vNM_N}(R)\, |M_S\rangle\ ,
\label{b7}
\end{align}
where,
\begin{equation}
V_{\textrm{SME}}^e(R;\boldsymbol{S}) ~=~ \langle N\, M_N |\, 
V_{\textrm{SME}}^e(\boldsymbol{R};\boldsymbol{S})\,|N\, M_N \rangle \ ,
\label{b8}
\end{equation}
together with,
\begin{equation}
\D E_{\rm SME}^n ~=~ \langle v\,N\,M_N|\,H_{\rm SME}^p\,|v\,N\,M_N \rangle \ .
~~~~~~~~~~~~~~~~~~~~~~
\label{b9}
\end{equation}

In the following two sections, we evaluate the SME contribution $V_{\rm SME}^e(R)$ to the inter-nucleon
potential in the hyperfine-Zeeman states and find the corresponding shifts in the rovibrational 
energy levels, then construct the direct proton contribution $\D E_{\rm SME}^n$.

\newpage

\section{Electron SME couplings and the inter-nucleon potential}\label{sect 3}

We now consider the construction of $V_{\rm SME}^e(R)$, the electron contribution to the 
inter-nucleon potential. (We suppress the explicit dependence on the spin label from now 
on for convenience and to match the notation in Paper 2.)
We start here with the spin-independent couplings $\VV_{njm}$, leaving the spin-dependent 
$\TT_{njm}^{(0B)}$, $\TT_{njm}^{(1B)}$ and $\TT_{njm}^{(1E)}$ to the following sub-section.

\subsection{Electron spin-independent SME couplings}\label {sect 3.1}

The results at $O({\bs p}^2)$ have already been derived using the couplings $c_{\m\n}$ and
$a_{\m\n\l}$ appearing in the original SME Lagrangian. Here, we instead start from the 
relativistic expansion of the Hamiltonian expressed in terms of spherical tensor couplings
and extend the discussion to include the leading terms arising at $O({\bs p}^4)$ in the
non-minimal SME. In this expansion, the SME Hamiltonian for the electron is written in terms
of non-relativistic, spin-independent couplings $\VV_{njm}^{e\,\NR}$ as
\begin{equation}
H_{\rm SME}^{e\,\VV} ~=~ - \sum_{njm} \,p^n Y_{jm}(\hat{\bs{p}})\,\VV_{njm}^{e\,\NR} \ ,
\label{b11}
\end{equation}
with $n$ summed over $0, 2, 4 \ldots$.  The couplings $\VV_{njm}^{e\,\NR}$ are related in a non-trivial
way to the original $\VV_{njm}^e$, the relations being given explicitly in \cite{Kostelecky:2013rta}.
These are especially important below when we consider various critical identities amongst the 
non-relativistic couplings.
Also recall that the couplings may be separated into \textsf{CPT} even ($c_{njm}^{e\,\NR}$)
and odd $(a_{njm}^{e\,\NR}$) components as follows:
\begin{equation}
\VV_{njm}^{e\,\NR} ~=~ c_{njm}^{e\,\NR} \,-\, a_{njm}^{e\,\NR} \ .
\label{b12}
\end{equation}

In this form, the Hamiltonian $H_{\rm SME}^e$ is independent of the frame, so we may consider it first 
in the \textsf{MOL} frame, denoting the relevant couplings by $\hat{\VV}_{njm}^{e\,\NR}$. 
The momentum expectation values are then evaluated as in Paper 1 with the familiar $1s\s_g$ electron
wavefunctions. 

It is very convenient in working with the spherical tensors to introduce the Wigner matrices
$d_{m' m}^j(\theta)$, in terms of which the spin-weighted spherical harmonics are written as
\begin{equation}
{}_sY_{jm}(\theta,\phi) ~=~ \sqrt{\frac{2j+1}{4\pi}}\, d_{m s}^j(\theta) e^{i m \phi} \ ,
\label{b13}
\end{equation}
with the usual spherical harmonics corresponding to $s=0$. The necessary properties of the
$d_{m' m}^j(\theta)$ in our conventions are given in section 6 and appendix A of Paper 2.

Using the explicit expressions for $d_{m0}^2$ and $d_{m 0}^4$, and imposing cylindrical symmetry on
the momentum expectation values, we find in the \textsf{MOL} frame,
\begin{align}
&V_{\rm SME}^{e\,\VV} ({\bs R}) ~=~ - \frac{1}{\sqrt{4\pi}} \bigg[ \hat{\VV}_{000}^{e\,\NR} 
\,+\, \langle p^2\rangle \, \hat{\VV}_{200}^{e\,\NR} \,-\, 
\frac{1}{2}\sqrt{5} \,\langle (p^2 - 3 p_z^2)\rangle \, \hat{\VV}_{220}^{e\,\NR} \nonumber \\
&~~~~+\, \langle p^4\rangle \,\hat{\VV}_{400}^{e\,\NR} \,-\, 
\frac{1}{2}\sqrt{5} \,\langle (p^4 - 3p^2 p_z^2)\rangle \,\hat{\VV}_{420}^{e\,\NR} 
\,+\, \frac{3}{8} \,\langle ( 3 p^4 - 30 p^2 p_z^2 + 35 p_z^4) \rangle \,\hat{\VV}_{440}^{e\,\NR} \,  +\,\ldots \bigg] 
\label{b14}
\end{align}
Note that in our previous work, we generally used the notation 
$\langle p^2\rangle = \tr\langle p_a \,p_b\rangle$ and
$\langle (p^2 - 3 p_z^2)\rangle = \tr_Y \langle p_a\,p_b\rangle$.

The cylindrical symmetry of the molecule enforces $m = 0$ here, since the momentum expectation values 
corresponding to $m\neq 0$ necessarily vanish. However, the absence of spherical symmetry does allow
$V_{\rm SME}^{e\,\VV} ({\bs R})$ to have contributions from the couplings with $j\neq 0$, to this order
$\hat{\VV}_{220}^{e\,\NR}$, $\hat{\VV}_{420}^{e\,\NR} $ and $\hat{\VV}_{440}^{e\,\NR}$. 

In our previous work (Paper 1), we have evaluated $ \tr \la p_a\,p_b \ra$ and $\tr_Y \la p_a\,p_b\ra$
as functions of $R$ and found their derivatives at the minimum $R_0$ of the inter-nucleon potential,
as required to evaluate the rovibrational energies in the form (\ref{aE}).
The non-zero value of $\tr_Y \la p_a\,p_b\ra$ arises because the molecule only has cylindrical and
not full spherical symmetry. It is smaller, but of the same magnitude as $\tr \la p_a\,p_b \ra$ and cannot
be neglected. For example, evaluated at $R_0$,  $\tr \la p_a\,p_b \ra =1.173$ while  $\tr_Y \la p_a\,p_b \ra = 0.360$
in units of the Bohr radius squared.

In the next step we rewrite the SME couplings in the \textsf{EXP} frame, with the quantisation ($z$) axis aligned
with the background magnetic field ${\bs B}$.  The transformation is given in terms of the Wigner matrices
by 
\begin{equation}
\hat{\VV}_{njm'} ~=~ \hat{\VV}_{njm}\, d_{m m'}^j(\theta)\, e^{i m \phi} \ ,
\label{b15}
\end{equation}
where $(\theta,\phi)$ are the standard spherical polar angles specifying the orientation of the \textsf{MOL} axes
in the \textsf{EXP} frame.

It follows immediately that
\begin{align}
V_{\rm SME}^{e\,\VV} ({\bs R}) ~=~ &- \bigg[ \frac{1}{\sqrt{4\pi}} \, \VV_{000}^{e\,\NR} 
\,+\, \frac{1}{\sqrt{4\pi}} \, \tr \la p_a\,p_b\ra\, \VV_{200}^{e\,\NR}   
\,-\, \frac{1}{2} \tr_Y\la p_a\,p_b\ra\, \VV_{22m}^{e\,\NR} \, Y_{2m}(\theta,\phi)  \nonumber \\
&~~~~~+ \frac{1}{\sqrt{4\pi}}\, \la p^4\ra \, \VV_{400}^{e\,\NR}   
\,-\, \frac{1}{2} \, \la (p^4 - 3 p^2 p_z^2)\ra \,\VV_{42m}^{e\,\NR}  \, Y_{2m}(\theta,\phi) \nonumber \\
&~~~~~+\, \frac{1}{8} \,\la (3 p^4 - 30 p^2 p_z^2 + 35 p_z^4) \ra \,
\VV_{44m}^{e\,\NR}  \, Y_{4m}(\theta,\phi) \,\,\bigg] \ .
\label{b16}
\end{align}
Note that $\VV_{000}^{e\,\NR}$ is usually neglected on the grounds that it is unobservable.

Following the discussion above, the next step is to take the expectation values of the spherical
harmonics in the angular momentum states $|N\, M_N\ra$ to find $V_{\rm SME}^{e\,\VV}(R)$ from (\ref{b8}).
In terms of Clebsch-Gordan coefficients,
\begin{equation}
\la N\, M_N| Y_{2m}(\theta,\phi) |N\,M_N\ra ~=~ \sqrt{\frac{5}{4\pi}}\, C_{N M_N, 2 0}^{NM_N}\, 
C_{N 0,2 0}^{N 0} \, \d_{m0} ~\equiv~ \sqrt{\frac{5}{4\pi}}\,c_{N M_N} \ ,
\label{b17}
\end{equation}
where,
\begin{equation}
c_{N M_N} ~=~ \frac{N(N+1) - 3M_N^2}{(2N-1)(2N+3)} \ .
\label{b18}
\end{equation}
This notation was used in our previous work. Here, we also need the extension to $j=4$,
\begin{equation}
\la N\, M_N| Y_{4m}(\theta,\phi) |N\,M_N\ra ~=~ \frac{3}{\sqrt{4\pi}}\, C_{N M_N, 4 0}^{NM_N}\, 
C_{N 0,4 0}^{N 0} \, \d_{m0} ~\equiv~  \frac{\,\,3}{\sqrt{4\pi}}\, c_{N M_N}^{(4)} \ ,
\label{b19}
\end{equation}
with 
\begin{equation}
c_{NM_N}^{(4)} ~=~ \frac{3}{4} \,\,\frac{\big[3N(N+2)(N+1)(N-1) \,-\,5 (6N^2 + 6N -5) M_N^2 \,+\, 35 M_N^4 \big]}
 {\big[(2N+5)(2N+3)(2N-1)(2N-3)\big]}\ .
\label{b20}
\end{equation}
The special case $N=2$ is of current experimental interest, so we note here the special cases\footnote{These 
expressions appear in \cite{Vargas:2025efi}, eq.(27), in the discussion of proton couplings.}
\begin{equation}
c_{2M_N} ~=~ \frac{1}{7} \big(2 - M_N^2\big) \ ,
\label{b21}
\end{equation}
and
\begin{equation}
c_{2  M_N}^{(4)} ~=~ \frac{1}{252} \big(72 \,-\,155M_N^2\,+\, 35 M_N^4 \big) \ .
\label{b22}
\end{equation}

These results are directly applicable in the strong magnetic field
limit, where $|v\,N\,M_J\,M_S\ra$ are the appropriate ``Zeeman'' eigenstates. 
Recall that $M_J = M_N + M_S$ is a good quantum number for all magnetic fields as the operator
$J_z$ commutes with the full hyperfine-Zeeman Hamiltonian.
This is the case of most interest for \Hbarmol where it
is envisaged spectroscopy will be carried out in Penning traps \cite{SchillerCERN2026}. 
In this limit, we find $V_{\rm SME}^{e\,\VV}(R)$
evaluated in the states $M_J, \,M_S = \pm\thalf$ is
\begin{align}
&V_{\rm SME\pm}^{e\,\VV} (R) ~=~ -\frac{1}{\sqrt{4\pi}}\bigg[ \VV_{000}^{e\,\NR} \,+\, 
\tr \la p_a\,p_b\ra \,\VV_{200}^{e\,\NR} \,-\, 
\frac{\sqrt{5}}{\,2} \,\tr_Y\la p_a\,p_b\ra \,c_{NM_J\mp\thalf}\,\VV_{220}^{e\,\NR}
\nonumber \\
&~~~~~~~~~~~~~~~~~~~~~~~~~~~~+~\la p^4\ra \, \VV_{400}^{e\,\NR} \,-\, 
\frac{\sqrt{5}}{\,2} \,\la\,(p^4 - 3 p^2 p_z^2)\,\ra\,c_{NM_J\mp\thalf}\, \VV_{420}^{e\,\NR} \nonumber \\
&~~~~~~~~~~~~~~~~~~~~~~~~~~~~~~~~~~~~~~~~~~~~~~+\,
\frac{3}{8} \,\la\,(3p^4 - 30 p^2 p_z^2 + 35 p_z^4)\,\ra \, c_{NM_J\mp\thalf}^{(4)} \, \VV_{440}^{e\,\NR}\, \bigg] 
\label{b22a}
\end{align}
For large, but finite, $B$ fields, the coefficients acquire corrections, but only of $O(1/B)^2$; these are calculated in 
Appendix \ref{appA}.

On the other hand, precision spectroscopy of \Hmol is already being carried out in RF traps at extremely
small magnetic fields \cite{SAS2024} where, as discussed in Paper 2, 
the relevant ``hyperfine'' eigenstates are $|v\,N\,J\,M_J\ra$
where ${\bs J} = {\bs N} + {\bs S}$ is the total molecular angular momentum.
Here, 
\beq
|v\,N\,J\,M_J\ra ~=~ \sum_{M_S}\, C_{N M_N, \tfrac{1}{2} M_S}^{J M_J} \, |v\, N\, M_N\,M_S\ra \ ,
\label{b23}
\eeq
with $J = N \pm \thalf$ and $M_J = -J, \ldots J-1, J$ as usual. 
We therefore need,
\beq
V_{\rm SME}^{e\,\VV} (R)  ~=~  \la v\,N\,J'\,M_J|\,V_{\rm SME}^{e\,\VV} ({\bs R})\,|v\,N\,J\,M_J\ra \ ,
\label{b24}
\eeq
where, for each $M_J$, this is a 2 x 2 matrix with rows/columns signifying $J', J = N\pm\thalf$.

At this point, we refer to Paper 2 and Appendix \ref{appA}, which describe the hyperfine-Zeeman Hamiltonian 
and its diagonalisation for non-vanishing magnetic fields, in order to find the SME perturbed energy eigenvalues.
The eigenvalues may be written as
\begin{align}
&V_{\rm SME\pm}^{e\,\VV} (R) ~=~ -\frac{1}{\sqrt{4\pi}}\bigg[ \VV_{000}^{e\,\NR} \,+\, 
\tr \la p_a\,p_b\ra \,\VV_{200}^{e\,\NR} \,-\, 
\frac{\sqrt{5}}{\,2} \,\tr_Y\la p_a\,p_b\ra \,\hat{c}_{NM_J}^\pm(B)\,\VV_{220}^{e\,\NR}
\nonumber \\
&~~~~~~~~~~~~~~~~~~~~~~~~~~~~+~\la p^4\ra \, \VV_{400}^{e\,\NR} \,-\, 
\frac{\sqrt{5}}{\,2} \,\la\,(p^4 - 3 p^2 p_z^2)\,\ra\,\hat{c}_{NM_J}^\pm(B) \, \VV_{420}^{e\,\NR} \nonumber \\
&~~~~~~~~~~~~~~~~~~~~~~~~~~~~~~~~~~~~~~~~~~~~~~+\,
\frac{3}{8} \,\la\,(3p^4 - 30 p^2 p_z^2 + 35 p_z^4)\,\ra \, \hat{c}_{NM_J}^{(4)}(B) \, \VV_{440}^{e\,\NR}\, \bigg] 
\label{b25}
\end{align}
with the coefficients,
\begin{align}
\wh{c}_{NM_J}^{\,+}(B) ~&=~ \frac{1}{(2N+1)(2N+3)}\,\big[(N+\thalf)(N+\tfrac{3}{2}) - 3 M_J^2\big]   
\nonumber \\[8pt]
&~~~~~~~~~~~~~~~~~~~~~~~~~~~~~~ 
+~24 \,\frac{\big[(N+\thalf)^2 - M_J^2\big]\, M_J}{(2N+1)^3 (2N-1)(2N+3)}\,\, {\cal B}  \ ,
\label{b26a}
\end{align}
and 
\begin{align}
\wh{c}_{NM_J}^{\,-}(B) ~&=~ \frac{1}{(2N-1)(2N+1)}\,\big[(N-\thalf)(N+\thalf) - 3 M_J^2\big] 
\nonumber \\[8pt]
&~~~~~~~~~~~~~~~~~~~~~~~~~~~~~~
 -~24 \,\frac{\big[(N+\thalf)^2 - M_J^2\big]\, M_J}{(2N+1)^3 (2N-1) (2N+3)}\,\, {\cal B}  \ .
\label{b26b} 
\end{align}
Here, we have introduced the notation
\begin{equation}
{\cal B}~=~ \frac{1}{c_e(v,N)}\, \big(g_e + g_m(v,N)\big) \mu_B B 
\label{b26c}
\end{equation}
for the expansion parameter, where $c_e(v,N),\,g_e,\, g_m(v,N)$ are coefficients in the 
hyperfine-Zeeman Hamiltonian.

At $O(p^4)$, citing only the terms independent of $B$ for simplicity,
we find
\begin{equation}
\hat{c}_{NM_J}^{(4)+} ~=~ \frac{3}{64}\,\, \frac{3(2N+5) (2N+3) (2N+1) (2N-1) - 40 (12N^2 + 24N -1) M_J^2 + 560 M_J^4}
{(2N+5)(2N+3)(2N-1)(2N-3)}  
\label{b27a}
\end{equation}
and
\begin{equation}
\hat{c}_{NM_J}^{(4)-} ~=~ \frac{3}{64}\, \,\frac{3(2N+3)(2N+1)(2N-1)(2N-3) - 40 (12N^2 - 13) M_J^2 + 560 M_J^4}
{(2N+3)(2N+1)(2N-1)(2N-3)} ~~~~\ ,
\label{b27b}
\end{equation}
using results from Appendix \ref{appB}.

In the presently relevant experimental case with $N=2$, these
expressions simplify at zero magnetic field to\footnote{For $N=2$, these factors 
have been given in \cite{Vargas:2025efi}, eq.(11).
To compare, the dictionary for our notations is:
\begin{equation*}
\a_{\frac{3}{2}M_F}^{(2)} ~=~ \sqrt{\frac{5}{4\pi}}\, \hat{c}_{2M_J}^- \ ,
~~~~~~~~
\a_{\frac{5}{2}M_F}^{(2)} ~=~ \sqrt{\frac{5}{4\pi}}\, \hat{c}_{2M_J}^+ \ ,
~~~~~~~~
\a_{\frac{3}{2}M_F}^{(4)} ~=~ \frac{3}{\sqrt{4\pi}}\, \hat{c}_{2M_J}^{(4)-} \ ,
~~~~~~~~~
\a_{\frac{5}{2}M_F}^{(4)} ~=~ \frac{3}{\sqrt{4\pi}}\, \hat{c}_{2M_J}^{(4)+} \ .
\end{equation*}
Note that $\hat{c}_{2M_J}^{(4)-}$ vanishes for both allowed values $M_J = 1/2$ or $3/2$.
}
\begin{align}
\hat{c}_{2M_J}^+ ~&=~ \frac{1}{140}\, \big(35  \,-\,12M_J^2\big)  \nonumber \\[5pt]
\hat{c}_{2M_J}^- ~&=~ \frac{1}{20}\, \big(5 \,-\, 4M_J^2\big)  \ ,
\label{b28}
\end{align}
and
\begin{align}
\hat{c}_{2M_J}^{(4)+}~&=~ \frac{1}{432} \, \big( 567 \,-\, 760 M_J^2 \,+\, 112 M_J^4 \big)  \nonumber \\[5pt]
\hat{c}_{2M_J}^{(4)-}~&=~ \frac{1}{64} \, \big(9 \,-\, 40 M_J^2 \,+\, 16 M_J^4 \big)  \ .
\label{b29}
\end{align}

The formulae (\ref{b22a}) and (\ref{b25}) for $V_{\rm SME\,\pm}^{e\,\VV} (R)$ form the starting point to express 
the rovibrational energies as a function of its derivatives with respect to $R$, as shown in
Paper 1.  Keeping terms up to $O\l^2)$, we have the SME contributions,
\begin{align}
E^{e\,\VV}_{vNM_J\pm} ~&=~ V_{\rm SME\,\pm}^{\,e\,\VV} 
\,+\,\d_{\rm SME\,\pm}^{\,e\,\VV}\,(v+\thalf)\,\omega_0  \,-\,  x_{\rm SME\,\pm}^{\,e\,\VV}\,(v+\thalf)^2\,\omega_0
\nonumber \\[6pt]
&~~~~~~~~~~~~~~\,+\,B_{\rm SME\,\pm}^{\,e\,\VV}\,N(N+1)\,\omega_0 \,-\, 
\a_{\rm SME\,\pm}^{\,e\,\VV}\,(v+\thalf) N(N+1) \,\omega_0
\,+\, \ldots
\label{b30}
\end{align}
where (with all derivatives evaluated at the minimum $R_0$ of the original inter-nucleon potential $V_M(R)$),
\begin{equation}
\d_{\rm SME}^e \,=\, \frac{1}{2} \,\frac{1}{V_M^{''}}\, \Big[\, V_{\rm SME}^{e\,''}  \,-\, 
\frac{V_M^{'''}}{V_M^{''}}\,\, V_{\rm SME}^{e\,'}\Big] \ ,
~~~~~~~~~~~
B_{\rm SME}^e \,=\,  \l\, \,\frac{1}{V_M^{''}}\, \Big[\frac{1}{R_0}\,V_{\rm SME}^{e\,'} \,\Big]  \ ,
\label{b30}
\end{equation}
and similar, much lengthier, expressions for $\a_{\rm SME}^{\,e}$ and $x_{\rm SME}^{\,e}$
(see Paper 1).

To complete the determination of the contributions to the rovibrational energies from the SME,
we need the expectation values of the momenta in (\ref{b22a}), (\ref{b25}) and their $R$-derivatives. 
This calculation was carried out in detail for the terms of $O(p^2)$ in Paper 1, Appendix A, and numerical 
results were presented there. This can be repeated in exactly the same way for the $O(p^4)$ expectation
values. However, at this point in the experimental programme it is probably excessive to present these here,
so we content ourselves with noting that compared to the $O(p^2)$ expectation values, they will be
parametrically suppressed by $O(1/a_0^2) \,=\, O(\a^2 m_e^2)$, where $a_0$ is the Bohr radius.
Their importance therefore depends on the relative magnitude of the $n=4$ SME couplings 
$\VV_{4j0}^{e\,{\rm NR}}$ compared to the $n=2$ couplings $\VV_{2j0}^{e\,{\rm NR}}$ which in  this context
should be viewed as {\it a priori} unknown.

Showing just the terms up to $O(p^2)$, we find from (\ref{b25}) for weak magnetic fields,\footnote{
There is a significant sign error in the analogous expressions in Paper 1. This originates in
eq.(4.23) where the sign of the coefficient of the $\tr_Y \tilde{E}_{ij}^e$ term should be reversed.
This is propagated through section 6, especially (6.8) which should be compared with the corresponding
expressions here, and section 7, especially (7.1), where the same sign error affects the coefficient 
of all terms with $(c_{220}^{\NR\, e} - a_{220}^{\NR\,e})$.  This sign error is repeated in Paper 2 in the
numerical results in eqs.(7.6) and (7.10), although all algebraic expressions in this paper are correct.
These are now corrected in the arXiv versions of the papers.}
\begin{align}
V_{\rm SME\,\pm}^{\,e\,\VV}  ~&=~~\frac{1}{\sqrt{4\pi}}\,  \bigg[-\,\VV_{000}^{\,e\,\NR} 
\,-\,2.346\,m_e\,\VV_{200}^{\,e\,\NR} \,+\, 0.360  \, \sqrt{5}\,\,\hat{c}_{NM_J}^\pm(B)\, m_e\,\VV_{220}^{\,e\,\NR} 
\,+\, \ldots \bigg]
\nonumber \\[8pt]
\d_{\rm SME\,\pm}^{\,e\,\VV}  ~&=~~\frac{1}{\sqrt{4\pi}}\,\bigg[
3.009\,m_e\,\VV_{200}^{\,e\,\NR} \,-\, 0.817  \, \sqrt{5}\,\,\hat{c}_{NM_J}^\pm(B)\,m_e\, \VV_{220}^{\,e\,\NR} 
\,+\, \ldots \bigg]
\nonumber \\[8pt]
B_{\rm SME\,\pm}^{\,e\,\VV}  ~&=~~ \l\,\frac{1}{\sqrt{4\pi}}\,\bigg[
2.002\,m_e\,\VV_{200}^{\,e\,\NR} \,-\, 0.336  \, \sqrt{5}\,\,\hat{c}_{NM_J}^\pm(B)\,m_e\, \VV_{220}^{\,e\,\NR} 
\,+\, \ldots \bigg]
\nonumber \\[8pt]
x_{\rm SME\,\pm}^{\,e\,\VV}  ~&=~~ \l\,\frac{1}{\sqrt{4\pi}}\,\bigg[ 
4.887\,m_e\,\VV_{200}^{\,e\,\NR} \,-\, 0.195  \, \sqrt{5}\,\,\hat{c}_{NM_J}^\pm(B)\,m_e\, \VV_{220}^{\,e\,\NR} 
\,+\, \ldots \bigg]
\nonumber \\[8pt]
\a_{\rm SME\,\pm}^{\,e\,\VV}  ~&=~~ \l^2\,\frac{1}{\sqrt{4\pi}}\,\bigg[ 
6.456\,m_e\,\VV_{200}^{\,e\,\NR} \,-\, 0.285  \, \sqrt{5}\,\,\hat{c}_{NM_J}^\pm(B)\,m_e\, \VV_{220}^{\,e\,\NR} 
\,+\, \ldots \bigg] \ ,
\label{b31}
\end{align}
with analogous results for strong fields from (\ref{b22a}) with 
$\hat{c}_{NM_J}^\pm \,\rta\, c_{N M_J \mp\thalf}^\pm$.
Note that $m_e \VV_{2jm}^{\,e\,\NR}$ and the coefficients $\d_{\rm SME\,\pm}^{\,e\,\VV}, \ldots$
are dimensionless. We are using atomic units here (see Paper 1) in which all energies are given in units of the
Rydberg energy $R_H$.

\subsection{Electron spin-dependent couplings}\label{sect 3.2}

We now come to the electron spin-dependent couplings. The non-relativistic SME Hamiltonian is
in this case \cite{Kostelecky:2013rta},
\begin{align}
H_{\rm SME}^{e\,\TT} ~=~ &- \s_{\HEL}^r \,\sum_{njm}\,p^n \,Y_{jm}(\hat{\bs p}) \,\TT_{njm}^{e\,\NR\,(0B)}  \nonumber \\
&+ \s_{\HEL}^{\pm} \,\sum_{njm}\, p^n\, {}_{\pm 1}Y_{jm}(\hat{\bs p})\,\big(
\pm \TT_{njm}^{e\,\NR\,(1B)} \,+\,i \TT_{njm}^{e\,\NR\,(1E)}\big) \ ,
\label{c1}
\end{align}
where the spins are given in the helicity frame (\textsf{HEL}) associated with the electron momentum.
The $\TT_{njm}^{\NR}$ are combinations of the equivalent \textsf{CPT} violating and conserving couplings,
\beq
\TT_{njm}^{\NR} ~=~ g_{njm}^{\NR} \,-\, H_{njm}^{\NR} \ ,
\label{c2}
\end{equation}
respectively, for each of the $(0B)$, $(1B)$ and $(1E)$ type couplings.
We only require \cite{Kostelecky:2013rta} those with $n=0,\,2,\,4\ldots$ and $j = (n+1), (n-1), \ldots \ge 0$
for the $(0B)$ couplings ($\ge 1$ for $(1B)$) and $j= n, (n-2), \ldots \ge 1$ for the $(1E)$ type.

Now, while the couplings in the full relativistic theory are independent, those in the non-relativistic approximation 
(\ref{c1}) are not, but rather satisfy some identities which will be important in our analysis.
Specifically, we can show\footnote{From the relations (112) in \cite{Kostelecky:2013rta}, we can 
re-express the following special cases of the non-relativistic couplings $g_{njm}^{\NR}$ in terms of the 
fully relativistic $g_{njm}^{(d)}$ as follows ($d$ is the dimension of the operator),
\begin{align*}
g_{01m}^{\NR\,(0B)} ~=~ \sum_{d,even}\,m^{d-3}\, g_{01m}^{(d)\,(0B)} ~=~ g_{01m}^{\NR\,(1B)}  \ , \nonumber \\
g_{23m}^{\NR\,(0B)} ~=~ 3 \sum_{d,even}\,m^{d-5}\, g_{23m}^{(d)\,(0B)} ~=~ \sqrt{\frac{3}{2}}\,\,g_{23m}^{\NR\,(1B)} \ ,
\end{align*}
after a short calculation, reflecting the fact that the putative couplings $g_{01m}^{(d)\,(1B)}$ 
and $g_{23m}^{(d)\,(1B)}$ are not permitted by the restrictions $n\ge 2$, $j=(n-1), (n-3), \ldots \ge1$ on the 
relativistic couplings $g_{njm}^{(d)\,(1B)}$.  See \cite{Kostelecky:2013rta}, Table III. 
Similar results hold for the $H_{njm}^{\NR}$.}
\begin{equation}
\TT_{01m}^{\NR\,(1B)} ~=~ \TT_{01m}^{\NR\,(0B)} \ ,
\label{c3}
\end{equation}
and 
\begin{equation}
\TT_{23m}^{\NR\,(1B)} ~=~ \sqrt{\frac{2}{3}}\, \TT_{23m}^{\NR\,(0B)} \ .
\label{c4}
\end{equation}
Note that these identities are not restricted to the minimal SME. 

The first step is to rewrite the Hamiltonian (\ref{c1}) entirely in the \MOL frame.
This means transforming the spin variables from the original helicity frame as presented in \cite{Kostelecky:2013rta}.
The transformations are given in terms of the Wigner matrices by
\begin{equation}
\s_{s'}^\HEL ~=~ \hat{\s}_s\, d_{s s'}^1(\theta) \, e^{-i s\phi} \ ,
\label{c5}
\end{equation}
where $\hat{}$ denotes the \textsf{MOL} frame variables and 
$(\theta,\phi)$ the direction of the electron momentum in the \textsf{MOL} frame.
Together with the definition (\ref{b13}) of the spin-weighted spherical harmonics, we can rewrite (\ref{c1})
in the \MOL frame as
\begin{align}
H_{\rm SME}^{e\,\TT} ~=~ - \sum_{njm}\, p^n \,\sqrt{\frac{2j+1}{4\pi}}\, e^{i(m-s)\phi} \,\bigg[
&~\hat{\s}_s\, d_{s0}^1\, d_{m0}^j \, \hat{\TT}_{njm}^{e\,\NR\,(0B)} \nonumber \\
&+\, \hat{\s}_s \,\big(d_{s,\,-1}^1\, d_{m,\,-1}^j \, +\, d_{s\,1}^1\, d_{m\,1}^j \big)\,\hat{\TT}_{njm}^{e\,\NR\,(1B)}
\nonumber \\
&+\, \hat{\s}_s \,\big(d_{s,\,-1}^1\, d_{m,\,-1}^j \, -\, d_{s\,1}^1\, d_{m\,1}^j \big)\,i\hat{\TT}_{njm}^{e\,\NR\,(1E)} ~ \bigg] \ .
\label{c6}
\end{align}

This expression is simplified using special cases of the following relations for products
of Wigner matrices,
\begin{equation}
d_{s'\,m'}^{j'}(\theta) \, d_{s\,m}^j(\theta) ~=~ \sum_{J=|j-1|}^{j+1}\,\,\, C_{j'\,s',\,j\,s}^{J\,s'+s}\,
C_{j'\,m',\,j\,m}^{J\,m'+m}\,\, d_{s'+s,\,m'+m}^J(\theta) \ ,
\label{c7}
\end{equation}
and exploiting the useful relations 
\begin{equation}
d_{m'\,m}^j ~=~ (-1)^{m'-m}\,d_{m\,m'}^j ~=~ d_{-m\,-m'}^j \ .
\label{c8}
\end{equation}

The calculation is described in detail for the minimal SME in Appendix A of Paper 2 and will not be 
repeated here.\footnote{The calculation of the $\hat{\TT}_{22m}^{e\,\NR\,(1E)}$ coupling presented in
Paper 2 is unnecessarily complicated and is greatly simplified by using the identity $d_{s,\,1}^1 \,=\, 
(-1)^{s-1}\,d_{-s,\,-1}^1$ before applying (\ref{c7}).} 
Also, to keep the presentation relatively simple, we only quote
results here up to $O(p^2)$ for the spin-dependent contributions.
We find, including the non-minimal $\hat{\TT}_{23m}^{e\,\NR}$ couplings but {\it without} at this stage 
applying the identities (\ref{c3}) and (\ref{c4}),
\begin{align}
H_{\rm SME}^{e\,\TT} ~=~ &~~Y_{00}(\theta,\phi)\,\hat{\s}_s\, (-1)^s \,C_{1\,-s,\,1\,m}^{0\,0}\, 
\big(\hat{\TT}_{01m}^{e\,\NR\,(0B)} \,+\,2\,\hat{\TT}_{01m}^{e\,\NR\,(1B)} \big) \nonumber  \\[3pt]
&- \sqrt{\frac{2}{5}}\, Y_{20}(\theta,\phi)\, \hat{\s}_s\,(-1)^s\, C_{1\,-s,\,1\,m}^{2\,0}\, 
\big(\hat{\TT}_{01m}^{e\,\NR\,(0B)} \,-\,\hat{\TT}_{01m}^{e\,\NR\,(1B)} \big) \nonumber  \\[3pt]
&+\, p^2\, Y_{00}(\theta,\phi)\,\hat{\s}_s\, (-1)^s \,C_{1\,-s,\,1\,m}^{0\,0}\, 
\big(\hat{\TT}_{21m}^{e\,\NR\,(0B)} \,+\,2\,\hat{\TT}_{21m}^{e\,\NR\,(1B)} \big) \nonumber  \\[3pt]
&- \sqrt{\frac{2}{5}}\, p^2\, Y_{20}(\theta,\phi)\, \hat{\s}_s\,(-1)^s\, C_{1\,-s,\,1\,m}^{2\,0}\, 
\big(\hat{\TT}_{21m}^{e\,\NR\,(0B)} \,-\,\hat{\TT}_{21m}^{e\,\NR\,(1B)} \big) \nonumber  \\[3pt]
&- \sqrt{2}\,p^2\, Y_{20}(\theta,\phi)\, \hat{\s}_s\,(-1)^s\, C_{1\,-s,\,2\,m}^{2\,0}\,
 i\, \hat{\TT}_{22m}^{e\,\NR\,(1E)}  \nonumber  \\[3pt]
&+\sqrt{\frac{3}{5}}\,p^2\, Y_{20}(\theta,\phi)\, \hat{\s}_s\,(-1)^s\, C_{1\,-s,\,3\,m}^{2\,0}\, 
\big(\hat{\TT}_{23m}^{e\,\NR\,(0B)} \,+\, 2\sqrt{\frac{2}{3}}\,\hat{\TT}_{23m}^{e\,\NR\,(1B)} \big) \nonumber  \\[3pt]
&-\frac{2}{3}\,p^2 \,Y_{40}(\theta,\phi)\, \hat{\s}_s\,(-1)^s\, C_{1\,-s,\,3\,m}^{4\,0}\, 
\big(\hat{\TT}_{23m}^{e\,\NR\,(0B)} \,-\, \sqrt{\frac{3}{2}}\,\hat{\TT}_{23m}^{e\,\NR\,(1B)} \big)  \ .
\label{c9}
\end{align}
Notice the important simplification at this point that the `unnatural' terms proportional
just to $Y_{20}$ and $p^2 Y_{40}$ vanish by virtue of the identities (\ref{c3}) and (\ref{c4}).
This just leaves the familiar combinations $p^2 \,Y_{00} \,=\, \tfrac{1}{\sqrt{4\pi}}\,\tr\,p_a\,p_b$
and $p^2\,Y_{20}\,=\,-\tfrac{\sqrt{5}}{2}\,\frac{1}{\sqrt{4\pi}}\, \tr_Y\,p_a\,p_b$.

The next step is to write the remaining Clebsch-Gordan coefficients in more manageable form.
We need,\footnote{We use the notation from Paper 2 for $\textsf{T}_{sm}$ and $\textsf{Y}_{sm}$
and also define the new matrix $\textsf{X}_{sm}$ arising in the non-minimal sector as:
\begin{equation*}
\textsf{T}_{sm} ~=~  \begin{pmatrix}
1 &~0&~0\\ 0&~-1&~0 \\0&~0& \,\,0\\
\end{pmatrix} \ , ~~~~~~~~~~~~~
\textsf{Y}_{sm} ~=~  \begin{pmatrix}
1 &~0&~0\\ 0&~1&~0 \\0&~0& \,-2\\
\end{pmatrix} \ , ~~~~~~~~~~~~~
\textsf{X}_{sm} ~=~  \begin{pmatrix}
1 &~0&~0\\ 0&~1&~0 \\0&~0& \,\sqrt{\frac{3}{2}}\\
\end{pmatrix} \ .
\end{equation*}
}
\begin{align}
(-1)^s \,C_{1\,-s,\,1\,m}^{0\,0} ~&=~ -\frac{1}{\sqrt{3}} \,\d_{sm} \ , ~~~~~~~~~~~~~~
(-1)^s \,C_{1\,-s,\,1\,m}^{2\,0} ~=~ -\frac{1}{\sqrt{6}}\, \textsf{Y}_{sm} \ ,  \nonumber \\
(-1)^s \,C_{1\,-s,\,2\,m}^{2\,0} ~&=~ -\frac{1}{\sqrt{2}}\, \textsf{T}_{sm} \ , ~~~~~~~~~~~~~~
(-1)^s \,C_{1\,-s,\,3\,m}^{2\,0} ~=~ -\sqrt{\frac{2}{7}} \,\textsf{X}_{sm} \ .
\label{c10}
\end{align}
So finally, now applying the identities  (\ref{c3}) and (\ref{c4}), the spin-dependent non-minimal 
SME Hamiltonian can be written in the \textsf{MOL} frame up to $O(p^2)$ as\footnote{For $s=0$, the coefficient 
of $\hat{\TT}_{230}^{e\,\NR\,(0B)}$ has previously appeared in eq.(132) of \cite{Kostelecky:2015nma}.}
\begin{align}
H_{\rm SME}^{e\,\TT} = &- \frac{1}{\sqrt{4\pi}} \,\bigg[
\sqrt{3}\, \hat{\s}_s \,\d_{sm} \hat{\TT}_{01m}^{e\,\NR\,(0B)}
~+~
\tr\,p_a\,p_b  \,\bigg(\frac{1}{\sqrt{3}}\,\hat{\s}_s\,\d_{sm} \, \big(\hat{\TT}_{21m}^{e\,\NR\,(0B)} 
+\,2\,\hat{\TT}_{21m}^{e\,\NR\,(1B)} \big) \bigg)\nonumber \\[3pt]
&~~~~~~~~~~~+~\tr_Y\,p_a\,p_b \,\,\bigg(\frac{1}{2\sqrt{3}}\,\hat{\s}_s\, \textsf{Y}_{sm}\,
\big(\hat{\TT}_{21m}^{e\,\NR\,(0B)} \,-\,\hat{\TT}_{21m}^{e\,\NR\,(1B)} \big) \nonumber \\[3pt]
&~~~~~~~~~~~~~~~~~~~~~~~~~~~~~~~ ~-~
\sqrt{\frac{7}{6}}\, \hat{\s}_s\,\textsf{X}_{sm}\, \hat{\TT}_{23m}^{e\,\NR\,(0B)}  
~+~ \frac{\sqrt{5}}{2}\, \hat{\s}_s \, \textsf{T}_{sm} \, i\, \hat{\TT}_{22m}^{e\,\NR\,(1E)} \bigg)\,\bigg] \ .
\label{c11}
\end{align}

With the Hamiltonian specified in the \MOL frame, we can use the expectation values 
$\tr\la p_a\,p_b\ra$ and $\tr_Y\la p_a\,p_b\ra$ of the electron momenta evaluated in Paper 1, exploiting
the cylindrical symmetry of the \Hmol\, and \Hbarmol\,  molecular ions.  Then, we need
to express $H_{\rm SME}^{e\,\TT}$ in terms of the \EXP frame couplings.

The spins and couplings are transformed from the \MOL to \EXP frames by
\begin{align}
\hat{\s}_{s'} ~&=~ \s_s\, d_{s\,s'}^1(\theta) \, e^{-i\,s\,\phi} \ , \nonumber \\
\hat{\TT}_{njm'} ~&=~ \TT_{njm}\, d_{m\,m'}^j(\theta) \, e^{-i\,m\,\phi} \ ,
\label{c12} 
\end{align}
where now the angles $(\theta,\phi)$ describe the orientation of the molecular axis relative to
the \EXP frame, whose $z$-axis is aligned with the background magnetic field. Notice
especially the importance of including all the $\hat{\TT}_{njm}$ couplings here,
not simply those with $m=0$.

The techniques required for the conversion are similar to those described above for the 
spin-independent couplings and the derivation has been carried through in the minimal
SME sector in Paper 2. Here, we simply quote the result, as always interpreting the expectation
value of $H_{\rm SME}^{e\,\TT}$ in the electron $1s\s_g$ state as a contribution
$V_{\rm SME}^{e\,\TT}({\bs R})$ to the inter-nucleon potential:
\begin{align}
V_{\rm SME}^{e\,\TT}({\bs R}) ~&=~ -~\sqrt{\frac{3}{4\pi}} \, \Big[ \, \s_m\,\,\TT_{01m}^{e\,\NR\,(0B)} 
~+~ \frac{1}{3} \,\, \tr\langle \,p_a\,p_b\,\rangle\,\s_m\,\big(\TT_{21m}^{e\,\NR\,(0B)} \,+\,
2\,\TT_{21m}^{e\,\NR\,(1B)} \big)\,\Big]   \nonumber \\[6pt]
&+~\sqrt{\frac{1}{10}}\,\,\tr_Y\langle \,p_a\,p_b\,\rangle\,\, \s_s\, (-1)^s \sum_M\, C_{1\,-s,\,1\,m}^{2\,M}
\,Y_{2M}(\theta,\phi)\,\,\big(\TT_{21m}^{e\,\NR\,(0B)} \,-\, \TT_{21m}^{e\,\NR\,(1B)} \big) \nonumber \\[4pt]
&-~\sqrt{\frac{1}{2}} \,\, \tr_Y\langle \,p_a\,p_b\,\rangle \,
\s_s \,\,(-1)^s\,\sum_M\,C_{1\,-s,\,2\,m}^{2\,M} \, Y_{2M}(\theta,\phi)\,\, i\,\TT_{22m}^{e\,\NR\,(1E)} 
\nonumber \\[6pt]
&-~ \frac{1}{2\sqrt{5}} \frac{\,7}{\sqrt{6}} \, \,\tr_Y\langle \,p_a\,p_b\,\rangle \,
\s_s \, \,(-1)^s\,\sum_M\,C_{1\,-s,\,3\,m}^{2\,M} \, Y_{2M}(\theta,\phi)\,\, \TT_{23m}^{e\,\NR\,(0B)}  \ .
\label{c13}
\end{align}
We emphasise here how the sensitivity of the rovibrational energies to the non-minimal couplings 
$\TT_{230}^{e\,\NR\,(0B)}$ arises only with $\tr_Y\la p_a\,p_b\ra$, again reflecting the fact that
the molecular ion only has cylindrical symmetry. This explains their absence in the
corresponding energy formulae quoted in \cite{Vargas:2025efi}.

We can now impose the simplification discussed in Paper 2 that allows us to restrict to the
\EXP frame couplings with $m=0$ only. 
The argument is that the hyperfine-Zeeman interactions do not mix states with different values of $M_J$.
Any matrix element of $V_{\rm SME}^{e\,\TT}({\bs R})$ which is off-diagonal in  $M_J$ is therefore of 
$O({\rm SME})$ and therefore contributes to energy levels only at $O({\rm SME})^2$, which we neglect.
Imposing $\D M_J$ on the matrix elements of $V_{\rm SME}^{e\,\TT}({\bs R})$ implies the
constraint $M+s=0$ in (\ref{c13}), and since $M=-s+m$ this in turn requires $m=0$.

Inserting the required Clebsch-Gordan coefficients\footnote{For $M=-s$, the Clebsch-Gordan coefficients 
appearing in (\ref{c13}) are:
\begin{align*}
(-1)^s\,C_{1\,-s,\,1\,0}^{2\,-s} ~&=~ - \frac{1}{\sqrt{2}}, ~ -\frac{1}{\sqrt{2}}, ~~ \sqrt{\frac{2}{3}} \ ,
~~~~~~~~~~~~~~~~~
(-1)^s\,C_{1\,-s,\,2\,0}^{2\,-s} ~=~ - \frac{1}{\sqrt{2}}, ~ \frac{1}{\sqrt{2}}, ~~ 0\ ,
\nonumber \\
(-1)^s\,C_{1\,-s,\,3\,0}^{2\,-s} ~&=~ - \frac{1}{\sqrt{7}}, ~ -\frac{1}{\sqrt{7}}, ~~ \sqrt{\frac{3}{7}} \ ,
~~~~~~~~~~~~~~~~{\rm for}\,\,  s = 1,\,-1,\,0 \,\, {\rm  respectively.}
\end{align*}
} we find, effectively,
\begin{align}
&V_{\rm SME}^{e\,\TT}({\bs R}) \,=\, -~\sqrt{\frac{3}{4\pi}} \, \Big[ \, \s_0\,\,\TT_{010}^{e\,\NR\,(0B)} 
~+~ \frac{1}{3} \,\, \tr\langle \,p_a\,p_b\,\rangle\,\s_0\,\big(\TT_{210}^{e\,\NR\,(0B)} +
2\,\TT_{210}^{e\,\NR\,(1B)} \big)\,\Big]   \nonumber \\[6pt]
&+~\sqrt{\frac{1}{15}}\,\,\tr_Y\langle \,p_a\,p_b\,\rangle\,\, \s_0\,
\,Y_{20}(\theta,\phi)\,\,\big(\TT_{210}^{e\,\NR\,(0B)} - \TT_{210}^{e\,\NR\,(1B)} +\,
\frac{3}{2}\,\sqrt{\frac{7}{6}}\, \TT_{230}^{e\,\NR\,(0B)} \big) \nonumber \\[4pt]
&-~ \frac{1}{2\sqrt{5}}\,\tr_Y\la p_a\,p_b\ra\, \bigg[
\s_1\, Y_{2\,-1}(\theta,\phi) \big( \TT_{210}^{e\,\NR\,(0B)} - \TT_{210}^{e\,\NR\,(1B)} -
\sqrt{\frac{7}{6}}\, \TT_{230}^{e\,\NR\,(0B)} -\sqrt{5}\,i\, \TT_{220}^{e\,\NR\,(1E)}\big) \nonumber\\
&~~~~~~~~~~~~~~~~~~~~~
+~ \s_{-1}\, Y_{2\,1}(\theta,\phi) \big( \TT_{210}^{e\,\NR\,(0B)} - \TT_{210}^{e\,\NR\,(1B)} -
\sqrt{\frac{7}{6}}\, \TT_{230}^{e\,\NR\,(0B)} + \sqrt{5}\,i\, \TT_{220}^{e\,\NR\,(1E)}\big) \bigg] \ .
\label{c14}
\end{align}

The final step in calculating the required rovibrational energies in the weak magnetic field regime 
is to find $V_{\rm SME}^{e\,\TT}(R)$ by
taking the matrix elements of (\ref{c14}) in the hyperfine states $|v\,N\,J\,M_J\ra$ as
in (\ref{b24}). Since this now also involves the spin operators $\s_s$, the calculation is much more
intricate, but all the required steps have already been set out in Paper 2 (eqs.~(5.12) -- (5.17)) and 
we will only quote the result here. For a given value of $M_J$, the states with $J= N\pm \thalf$ are mixed
and $V_{\rm SME}^{e\,\TT}(R)$ is best presented as a $2\times 2$ matrix with rows/columns representing
$J',\,J = N\pm \thalf$,
\begin{equation}
V_{\rm SME}^{e\,\TT}(R) ~=~ \sqrt{\frac{3}{4\pi}}\, \begin{pmatrix}
\a &~\b\\ \b^*&~\d \\
\end{pmatrix}  \ ,
\label{c15}
\end{equation}
After some calculation, we find
\begin{align}
\a ~=~ &- 2\,\frac{1}{2N+1}\,M_J  \,\,\TT_{010}^{e\,\NR\,(0B)} \nonumber \\[6pt]
&- \, \tr\,\langle\,p_a\,p_b\,\rangle\,
\, \frac{2}{3} \,\frac{1}{2N+1}\,M_J    \,\,   \big(\TT_{210}^{e\,\NR\,(0B)} + 2\,\TT_{210}^{e\,\NR\,(1B)} \big) \nonumber \\[8pt]
&~ +  \, \tr_Y\langle\,p_a\,p_b\,\rangle\, \, \frac{1}{(2N+1)(2N-1)(2N+3)} \nonumber\\[8pt]
&~~~~~~~~~~~~~~~~ \times~\bigg[
-\frac{2}{3} \, N(2N-1) \, M_J\,\,\big(\TT_{210}^{e\,\NR\,(0B)} -\TT_{210}^{e\,\NR\,(1B)} \big)  \nonumber \\[4pt]
&~~~~~~~~~~~~~~~~~~~~~~\, +\, 
\Big(3N^2 +6N +\frac{5}{4} - 5M_J^2\Big)
\,M_J\,\,\sqrt{\frac{7}{6}}\,\,\TT_{230}^{e\,\NR\,(0B)}  \bigg]
\label{c16}
\end{align}
and 
\begin{align}
\d ~~=~~&\,2\,\frac{1}{2N+1}\,M_J  \,\,\TT_{010}^{e\,\NR\,(0B)} \nonumber \\[6pt]
&+\,\tr\,\langle\,p_a\,p_b\,\rangle\,\,\frac{2}{3} \,\frac{1}{2N+1}\,M_J 
\,\big(\TT_{210}^{e\,\NR\,(0B)} + 2\,\TT_{210}^{e\,\NR\,(1B)} \big) \nonumber \\[8pt]
&-  \, \tr_Y\langle\,p_a\,p_b\,\rangle\, \, \frac{1}{(2N+1)(2N-1)(2N+3)} \nonumber\\[8pt]
&~~~~~~~~~~~~~~~~ \times~ \bigg[
-\frac{2}{3} \, (N+1)(2N+3) \,M_J\,\,\big(\TT_{210}^{e\,\NR\,(0B)} -\TT_{210}^{e\,\NR\,(1B)} \big)  \nonumber \\[4pt]
&~~~~~~~~~~~~~~~~~~~~~~~ +\, 
\Big(3N^2 - \frac{7}{4} -5M_J^2\Big) 
\,M_J\,\, \sqrt{\frac{7}{6}}\,\,\TT_{230}^{e\,\NR\,(0B)}  \,\bigg]
\label{c17}
\end{align}
while,
\begin{align}
\Re\,\b ~=~ &\sqrt{(N+\thalf)^2 -M_J^2}\,\,\,\, \bigg(
2\,\frac{1}{2N+1}\,\, \TT_{010}^{e\,\NR\,(0B)} \nonumber \\[6pt]
&+\,\tr\,\langle\,p_a\,p_b\,\rangle\, \,\frac{2}{3}\,\frac{1}{2N+1}\,\,
\big(\TT_{210}^{e\,\NR\,(0B)} + 2\,\TT_{210}^{e\,\NR\,(1B)} \big) 
\nonumber \\[8pt]
& +\, \tr_Y\langle\,p_a\,p_b\,\rangle\, \, \frac{1}{(2N+1)(2N-1)(2N+3)} 
\nonumber\\[8pt]
&\times~\bigg[
-\frac{1}{6}\,(2N-1)(2N+3)\,\,\big(\TT_{210}^{e\,\NR\,(0B)} -\TT_{210}^{e\,\NR\,(1B)} \big)  \nonumber \\[6pt]
&~~~~~~~~ -\, 
\Big(N^2 + N - \frac{3}{4} -5M_J^2\Big) \,\,
\sqrt{\frac{7}{6}}\,\,\TT_{230}^{e\,\NR\,(0B)}  \bigg]\,\bigg)\ .
\label{c18}
\end{align}

The coupling $\TT_{230}^{e\,\NR\,(1E)}$ drops out from $\Re\,\b$.  This leaves just 3 independent SME couplings
contributing to $V_{\rm SME}^{e\,\TT}(R)$ at $O(p^2)$, including the non-minimal coupling $\TT_{230}^{e\,\NR\,(0B)}$,
in line with the analysis in Paper 2 using the Lorentz tensor couplings of (\ref{a1}).

We can now summarise our results for the eigenvalues $V_{\rm SME\,\pm}^{e\,\TT}(R)$ in a concise form
suitable for reading off the contributions of the spin-dependent SME couplings to the rovibrational  energies.
For weak magnetic fields, applying the methods of Appendix \ref{appA}, and using the notation (\ref{b26c})
for the expansion parameter, we find:\footnote{In this paper,
we are using a self-consistent notation where hatted coefficients refer to the weak magnetic field regime
with states $|v\,N\,J\,M_J\ra$, while unhatted coefficients refer to the strong field regime
with states $|v\,N\,M_J\,M_S\ra$.  This means that the hatted $\hat{f}_{NM_J}$ coefficients in 
(\ref{c20b}), (\ref{c21a}) and (\ref{c21b}) here are identical to the corresponding unhatted $f_{NM_J}$ in Paper 2.}
\begin{align}
V_{\rm SME\,\pm}^{e\,\TT}(R) &= \sqrt{\frac{3}{4\pi}}\,\bigg(\hat{e}_{NM_J}^{\pm}(B)\,\, \TT_{010}^{e\,\NR\,(0B)}  
- \,\frac{1}{3}\,\tr\la p_a\,p_b\ra\,\, \hat{f}_{NM_J}^{\,\pm}(B)\,\big(\TT_{210}^{e\,\NR\,(0B)} + 
2\, \TT_{210}^{e\,\NR\,(1B)} \big)
\nonumber \\[6pt]
&~~~~~~~~~~~~~~~  - \,\frac{1}{3}\,\tr_Y\la p_a\,p_b\ra \,\, \hat{f}_{NM_J}^{\,Y\,\pm}(B)\, 
\big(\TT_{210}^{e\,\NR\,(0B)} - \TT_{210}^{e\,\NR\,(1B)} \big)
\nonumber \\[6pt]
&~~~~~~~~~~~~~~~ +\sqrt{\frac{7}{6}}\,\,\tr_Y\la p_a\,p_b\ra\,\, \hat{f}_{NM_J}^{\,X\,\pm}(B) 
\, \TT_{230}^{e\,\NR\,(0B)} \,\bigg) \ ,
\label{c19}
\end{align}
where
\begin{align}
\hat{e}_{NM_J}^{\,\pm}(B) ~&=~ \mp \frac{2}{2N+1}\,M_J  \,\mp\, \frac{8}{(2N+1)^3} \,[(N+\thalf)^2 -M_J^2]
\,\,\, {\cal B} \ , 
\label{c20a} \\[10pt]
\hat{f}_{NM_J}^{\,\pm}(B) ~&=~  \pm\frac{2}{2N+1} \,M_J ~\pm~ \frac{8}{(2N+1)^3} \,\big[(N+\thalf)^2 - M_J^2\big]
\,\,\,\, {\cal B}  \ , 
\label{c20b}
\end{align}
while
\begin{align}
\hat{f}_{NM_J}^{\,Y\,+}(B) ~&= ~~\frac{2N}{(2N+1)(2N+3)} \,M_J  
~~-~ \frac{2}{(2N+1)^3} \,\big[(N+\thalf)^2 - M_J^2\big]
\,\,\, {\cal B} \ , 
\label{c21a} \\[10pt]
\hat{f}_{NM_J}^{\,Y\,-}(B) ~&= - \frac{2(N+1)}{(2N-1)(2N+1)} \,M_J 
~~+~ \frac{2}{(2N+1)^3} \,\big[(N+\thalf)^2 - M_J^2\big]
\,\,\, {\cal B} \ , 
\label{c21b}
\end{align}
and
\begin{align}
\hat{f}_{NM_J}^{\,X\,+}(B) ~&=~~\frac{1}{4}\,\frac{12N^2 + 24N +5 -20M_J^2}{(2N+1)(2N-1)(2N+3)}\,M_J \nonumber \\[10pt]
&~~~~~~~~~~~~
~+~ \frac{(2N-1)(2N+3) -20M_J^2}{(2N+1)^3 (2N-1) (2N+3)}\,
\,\big[(N+\thalf)^2 - M_J^2\big]
\,\,\, {\cal B}  \ , 
\label{c22a} \\[14pt]
\hat{f}_{NM_J}^{\,X\,-}(B) ~&=-\,\,\frac{1}{4}\,\frac{12N^2 - 7 - 20 M_J^2}{(2N+1)(2N-1)(2N+3)}\,M_J \nonumber \\[10pt]
&~~~~~~~~~~~~
~-~\frac{(2N-1)(2N+3) -20M_J^2}{(2N+1)^3 (2N-1) (2N+3)}\,
\,\big[(N+\thalf)^2 - M_J^2\big]
\,\,\, {\cal B} \ , 
\label{c22b}
\end{align}
Notice that in all cases, the $O(B)$ correction vanishes for the unmixed, or stretched, states 
with $M_J = \pm (N+\thalf)$.

\vskip0.3cm
For large magnetic fields, we repeat this analysis taking the matrix elements of $V_{\rm SME}^{e\,\TT}({\bs R})$
in (\ref{c14}) in the basis states $|v\,N\,M_J\, M_S\ra$.  Some intermediate details are given in Appendix \ref{appA}.
Here, we simply present the result for the eigenvalues $V_{\rm SME\,\pm}^{e\,\TT}(R)$. 
This takes the same form as above,
\begin{align}
V_{\rm SME\,\pm}^{e\,\TT}(R) ~&=~ \sqrt{\frac{3}{4\pi}}\,\bigg(\, e_{NM_J}^{\pm}\,\, \TT_{010}^{e\,\NR\,(0B)}  
~- \,\frac{1}{3}\,\tr\la p_a\,p_b\ra\,\, f_{NM_J}^{\,\pm}\,\big(\TT_{210}^{e\,\NR\,(0B)} + 
2\, \TT_{210}^{e\,\NR\,(1B)} \big)
\nonumber \\[6pt]
&~~~~~~~~~~~~~~~  - \,\frac{1}{3}\,\tr_Y\la p_a\,p_b\ra \,\, f_{NM_J}^{\,Y\,\pm}(B)\, 
\big(\TT_{210}^{e\,\NR\,(0B)} - \TT_{210}^{e\,\NR\,(1B)} \big)
\nonumber \\[6pt]
&~~~~~~~~~~~~~~~ +\sqrt{\frac{7}{6}}\,\,\tr_Y\la p_a\,p_b\ra\,\, f_{NM_J}^{\,X\,\pm}(B) \, \TT_{230}^{e\,\NR\,(0B)} \,\bigg) \ ,
\label{c23}
\end{align}
with coefficients,
\begin{align}
e_{NM_J}^{\pm} ~&=~ \mp 1  \ ,
\label{c24} \\[8pt]
f_{NM_J}^{\pm}~&=~ \pm 1  \ ,
\label{c25} \\[8pt]
f_{NM_J}^{\,Y\,\pm}(B) ~&=~ \mp\, c_{N M_J \mp \thalf} ~~\pm~\, 3 \, \frac{[(N+\thalf)^2 - M_J^2]}{(2N-1)(2N+3)}\,\,
M_J\,\,\frac{1}{{\cal B}} \ ,~~~~~~~~~~~~~~
\label{c26} \\[10pt]
f_{NM_J}^{\,X\,\pm}(B) ~&=~ \pm \frac{1}{2}\, c_{N M_J \mp \thalf} ~~\pm~\,  \frac{[(N+\thalf)^2 - M_J^2]}{(2N-1)(2N+3)}\,\,
M_J\,\,\frac{1}{{\cal B}} \ .
\label{c27}
\end{align}

These coefficients, together with their weak-field counterparts, in principle allow constraints on the 
individual components of the electron SME couplings $\TT_{njm}^{\,e\,\NR}$ to be distinguished by measuring
suitably chosen combinations of transition frequencies.

\vskip0.5cm
In the same way as for the spin-independent couplings, we can calculate the contribution of the
inter-nucleon potentials $V_{\rm SME\,\pm}^{e\,\TT}(R)$ in (\ref{c19}) and (\ref{c23}) to the rovibrational
energies in the form,
\begin{align}
E^{e\,\TT}_{vNM_J\pm} ~&=~ V_{\rm SME\,\pm}^{\,e\,\TT} 
\,+\,\d_{\rm SME\,\pm}^{\,e\,\TT}\,(v+\thalf)\,\omega_0  \,-\,  x_{\rm SME\,\pm}^{\,e\,\TT}\,(v+\thalf)^2\,\omega_0
\nonumber \\[6pt]
&~~~~~~~~~~~~~~\,+\,B_{\rm SME\,\pm}^{\,e\,\TT}\,N(N+1)\,\omega_0 \,-\, 
\a_{\rm SME\,\pm}^{\,e\,\TT}\,(v+\thalf) N(N+1) \,\omega_0
\,+\, \ldots \ .
\label{c28}
\end{align}
Here, for weak magnetic fields, we find from (\ref{c19}),
\begin{align}
V_{\rm SME\,\pm}^{e\,\TT} &= \sqrt{\frac{3}{4\pi}}\,\bigg( 
\frac{2}{\a^2}\,\hat{e}_{NM_J}^{\pm}\,\frac{1}{m_e}\TT_{010}^{e\,\NR\,(0B)}  
\,- \,0.782\, \hat{f}_{NM_J}^{\,\pm}\,m_e\big(\TT_{210}^{e\,\NR\,(0B)} + 2\, \TT_{210}^{e\,\NR\,(1B)} \big)
\nonumber \\[6pt]
&~~~~~~ ~- \,0.240\, \hat{f}_{NM_J}^{\,Y\,\pm}\, 
m_e\big(\TT_{210}^{e\,\NR\,(0B)} - \TT_{210}^{e\,\NR\,(1B)} \big)
\,+ \,0.294 \sqrt{7}\, \hat{f}_{NM_J}^{\,X\,\pm} \, m_e\TT_{230}^{e\,\NR\,(0B)} \,\bigg) \ ,
\end{align}
while
\begin{align}
\d_{\rm SME\,\pm}^{\,e\,\TT} &= \sqrt{\frac{3}{4\pi}}\,\bigg(
1.003\, \hat{f}_{NM_J}^{\,\pm}\,m_e\big(\TT_{210}^{e\,\NR\,(0B)} + 2\, \TT_{210}^{e\,\NR\,(1B)} \big)
\nonumber \\[6pt]
&~~~~~~~ + 0.545\, \hat{f}_{NM_J}^{\,Y\,\pm}\, 
m_e\big(\TT_{210}^{e\,\NR\,(0B)} - \TT_{210}^{e\,\NR\,(1B)} \big)
\,-\, 0.667 \sqrt{7}\, \hat{f}_{NM_J}^{\,X\,\pm} \, m_e\TT_{230}^{e\,\NR\,(0B)} \,\bigg) \ ,
\nonumber\\[10pt]
B_{\rm SME\,\pm}^{\,e\,\TT} &= \l\,\sqrt{\frac{3}{4\pi}}\,\bigg(
0.667\, \hat{f}_{NM_J}^{\,\pm}\,m_e\big(\TT_{210}^{e\,\NR\,(0B)} + 2\, \TT_{210}^{e\,\NR\,(1B)} \big)
\nonumber \\[6pt]
&~~~~~~~ + 0.224\, \hat{f}_{NM_J}^{\,Y\,\pm}\, 
m_e\big(\TT_{210}^{e\,\NR\,(0B)} - \TT_{210}^{e\,\NR\,(1B)} \big)
\,-\, 0.275 \sqrt{7}\, \hat{f}_{NM_J}^{\,X\,\pm} \, m_e\TT_{230}^{e\,\NR\,(0B)} \,\bigg) \ ,
\nonumber\\[10pt]
x_{\rm SME\,\pm}^{\,e\,\TT} &= \l\,\sqrt{\frac{3}{4\pi}}\,\bigg(
1.629\, \hat{f}_{NM_J}^{\,\pm}\,m_e\big(\TT_{210}^{e\,\NR\,(0B)} + 2\, \TT_{210}^{e\,\NR\,(1B)} \big)
\nonumber \\[6pt]
&~~~~~~~ + 1.130\, \hat{f}_{NM_J}^{\,Y\,\pm}\, 
m_e\big(\TT_{210}^{e\,\NR\,(0B)} - \TT_{210}^{e\,\NR\,(1B)} \big)
\,-\, 1.384 \sqrt{7}\, \hat{f}_{NM_J}^{\,X\,\pm} \, m_e\TT_{230}^{e\,\NR\,(0B)} \,\bigg) \ ,
\nonumber\\[10pt]
\a_{\rm SME\,\pm}^{\,e\,\TT} &= \l^2\,\sqrt{\frac{3}{4\pi}}\,\bigg(
2.152\, \hat{f}_{NM_J}^{\,\pm}\,m_e\big(\TT_{210}^{e\,\NR\,(0B)} + 2\, \TT_{210}^{e\,\NR\,(1B)} \big)
\nonumber \\[6pt]
&~~~~~~~ + 0.195\, \hat{f}_{NM_J}^{\,Y\,\pm}\, 
m_e\big(\TT_{210}^{e\,\NR\,(0B)} - \TT_{210}^{e\,\NR\,(1B)} \big)
\,-\, 0.239 \sqrt{7}\, \hat{f}_{NM_J}^{\,X\,\pm} \, m_e\TT_{230}^{e\,\NR\,(0B)} \,\bigg) \ ,
\label{c29}
\end{align}
all in atomic units, with $m_e \TT_{2jm}^{\,e\,\NR}$ dimensionless and all energies in units of $R_H$.
For clarity, we have omitted the explicit $B$-dependence in the notation for the coefficients here.

The analogous results for strong magnetic fields follow in the same way from (\ref{c23}),
with the substitution from hatted to unhatted coefficients, $\hat{f}_{NM_J}^{\,\pm} 
\,\rta\, f_{NM_J}^{\,\pm}$ \,{\it etc}.

\newpage

\section{Proton SME couplings and rovibrational energies}\label{sect 4}

We now come to the proton sector of the SME and calculate the dependence of the rovibrational energy 
levels on the couplings $\VV_{njm}^{\,p}$.  Since we only consider para-\Hmol where the total proton spin
${\bs I} = 0$, we may restrict here to the spin-independent proton SME couplings alone.

In the non-relativistic approximation, the SME Hamiltonian for the two protons,
with momenta ${\bs p}_1$ and ${\bs p}_2$, is
\begin{equation}
H_{\rm SME}^p ~=~ - \sum_{njm}\, p_1^n \,Y_{jm}(\hat{\bs p}_1)\,\VV_{njm}^{\,p\,\NR} 
~-~\sum_{njm}\, p_2^n \,Y_{jm}(\hat{\bs p}_2)\,\VV_{njm}^{\,p\,\NR} ~~~~
\label{d1}
\end{equation}
with the sum running over $n=0,2,4,\ldots$. This expression is frame-independent, and we 
immediately interpret it in the \EXP frame.

The momenta ${\bs p}_1$, ${\bs p}_2$ and the electron momentum ${\bs p_e}$ are most conveniently re-expressed
in terms of the relative momentum ${\bs P}$ of the two protons, the momentum ${\bs p}$ of the electron
relative to the molecule centre of mass (CM), and the CM momentum ${\bs P}_{CM}$ of the molecule.
The momentum ${\bs P}$ is defined with a reduced mass $\mu = m_p/2$ and ${\bs p}$ with $\hat{\mu}
= 2 m_p m_e/(2m_p +m_e)$. Details are given in Paper 1. Setting ${\bs P}_{CM}$ to zero, the relations we need 
here are simply,
\begin{equation}
{\bs p}_1 ~=~ {\bs P} \,-\, \frac{1}{2} \,{\bs p} \ , ~~~~~~~~~~~~~~~~~~~~~~
{\bs p}_2 ~=~ - {\bs P} \,-\, \frac{1}{2}\, {\bs p} \ .
\label{d2}
\end{equation}

The momentum-independent term in $H_{\rm SME}^p$ is evidently just the constant
\begin{equation}
H_{\rm SME}^p\big|_{n=0}  ~=~  - \frac{2}{\sqrt{4\pi}} \,\, \VV_{000}^{\,p\,\NR} \ ,
\label{d3}
\end{equation}
and is usually considered unobservable and therefore neglected.
At $O(p^2)$, recalling the restrictions $j=n, n-2, \ldots \ge 0$ on the spherical
tensors \cite{Kostelecky:2013rta}, 
\begin{equation}
H_{\rm SME}^p\big|_{n=2} ~=~ - (p_1^2 + p_2^2) \,\frac{1}{\sqrt{4\pi}}\,\, \VV_{200}^{\,p\,\NR}
~-~ \big( \,p_1^2 \,Y_{2m}(\hat{\bs p}_1) \,+\, p_2^2\, Y_{2m}(\hat{\bs p}_2) \,\big)\,\, \VV_{22m}^{\,p\,\NR}\ .
\label{d4}
\end{equation}

Now, from (\ref{d2}),
\begin{equation}
p_1^2\,+\,p_2^2 ~=~ 2\, P^2 \,+\, \frac{1}{2}\,p^2 \ ,
\label{d5}
\end{equation}
where, crucially, the mixed terms involving ${\bs P}\cdot{\bs p}$ cancel.
In the Born-Oppenheimer analysis, this means that, as already described in Paper 1, the electron
Schr\"odinger equation receives an extra contribution from the proton SME couplings. 

To recover the equivalent split for the $j=2$ couplings, we need the following key result:
\begin{equation}
\,p_1^2 \,Y_{2m}(\hat{\bs p}_1) \,+\, p_2^2\, Y_{2m}(\hat{\bs p}_2) ~=~
2 P^2 \,Y_{2m}(\widehat{\bs P}) \,+\, \frac{1}{2}\,p^2\, Y_{2m}(\hat{\bs p}) \ .
\label{d6}
\end{equation}
To establish this, we use the explicit form of the spherical harmonics for each $m$ to write the l.h.s.~in terms
of Cartesian components of ${\bs p}_1$ and ${\bs p}_2$, then reconstruct the r.h.s.~following their re-expression
in terms of ${\bs P}$ and ${\bs p}$ using (\ref{d2}).

At this point, however, the second term is still expressed in the \EXP frame. Using the transformation (\ref{b15})
of the spherical tensors under rotations together with the expression (\ref{b13}) of the spherical harmonics
in terms of Wigner matrices, we note that this term is invariant, so we can rewrite it in terms of
\MOL frame couplings and momenta:
\begin{equation}
p^2\, Y_{2m}(\hat{\bs p}_\EXP) \,\, \VV_{22m}^{\,p\,\NR} ~=~ 
p^2\, Y_{2m}(\hat{\bs p}_\MOL) \,\, \hat{\VV}_{22m}^{\,p\,\NR} \ .
\label{d7}
\end{equation}
In this form we can take the expectation values of the momentum factors in the $1s\s_g$ state,
where cylindrical symmetry restricts to $m=0$, leaving 
\begin{equation}
\la\, p^2\, Y_{20}(\hat{\bs p}_\MOL) \,\ra \,\, \hat{\VV}_{220}^{\,p\,\NR} ~=~
-\frac{1}{2} \sqrt{\frac{5}{4\pi}}\, \tr_Y\la p_a\,p_b\ra \,\, \hat{\VV}_{220}^{\,p\,\NR} \ .
\label{d8}
\end{equation}
The total contribution at $O(p^2)$ to the electron Schr\"odinger equation is therefore
\begin{equation}
-\frac{1}{2}\,\frac{1}{\sqrt{4\pi}} \,\Big( \,\tr \la p_a\,p_b\ra\,\,\hat{\VV}_{200}^{\,p\,\NR} ~-~ 
\frac{1}{2}\,\sqrt{5}\,\tr_Y\la p_a\,p_b\ra\,\,\hat{\VV}_{220}^{\,p\,\NR} \Big) \\ .
\label{d9}
\end{equation}
Comparing with (\ref{b14}), the effect of these terms on the rovibrational energy levels is therefore 
included simply by substituting $\VV_{2jm}^{\,e\,\NR}$ with
\begin{equation}
\tilde{\VV}_{2jm}^{\,e\,\NR} ~\equiv~ \VV_{2jm}^{\,e\,\NR} ~+~ \frac{1}{2}\, \VV_{2jm}^{\,p\,\NR} \ .
\label{d10}
\end{equation}
in (\ref{b22a}) and (\ref{b25}). This is the equivalent of the result initially found in terms of Lorentz tensor
SME couplings in Paper 1. (Notice the absence of explicit factors of $m_e$ and $m_p$ in (\ref{d10}), compatible
with the identifications of the \textsf{CPT} even and odd components $\tilde{c}_{2jm}^e$ and 
$\tilde{a}_{2jm}^e$ in section 7 there.)

The remaining terms in the proton SME Hamiltonian, expressed in the \EXP frame, are then simply
\begin{equation}
H_{\rm SME}^p\big|_{n=0,2} ~=~ - \frac{2}{\sqrt{4\pi}}\,\, \VV_{000}^{\,p\,\NR} ~-~
2\, P^2\,\Big( \frac{1}{\sqrt{4\pi}}\,\, \VV_{200}^{\,p\,\NR} ~+~ 
Y_{2m}(\widehat{\bs P})\,\, \VV_{22m}^{\,p\,\NR}\,\Big) \ ,
\label{d11}
\end{equation}
and we recall from the discussion in the previous section that only the $m=0$ components contribute 
at $O({\rm SME})$.
The corresponding rovibrational energies follow by taking the expectation values first in the Zeeman
states $|v\,N\,M_J\,M_S\ra$ appropriate for high magnetic fields and then in the hyperfine states
$|v\,N\,J\,M_J\ra$ relevant at weak fields. 

The expectation values of the spherical harmonic $Y_{20}(\widehat{\bs P})$ in these states 
are the same as already encountered for the electron couplings. In the Zeeman states,
\begin{equation}
\la v\,N\,M_N\,M_S^\prime|\,Y_{20}(\widehat{\bs P})\,|v\,N\,M_N\ M_S\ra ~=~ 
\sqrt{\frac{5}{4\pi}}\, c_{N M_N} \, \d_{M_S M_S^\prime} \ ,
\label{d13}
\end{equation}
and we find
\begin{equation} 
\D E_{\rm SME\,\pm}^n\big|_{n=0,2} ~=~ -\frac{2}{\sqrt{4\pi}}\,\, \VV_{000}^{\,p\,\NR} ~-~
2 \,\la v\, N|\,P^2\,|v\, N\ra\, \Big(\frac{1}{\sqrt{4\pi}}\,\,\VV_{200}^{\,p\,\NR}
~+~ \sqrt{\frac{5}{4\pi}}\,c_{N M_J\mp\thalf}\,\,\VV_{220}^{\,p\,\NR}\,\Big) \ ,
\label{d12}
\end{equation}
up to $O(P^2)$, with the $c_{NM_N}$ coefficients in (\ref{b18}).

For the hyperfine states, the expectation values at fixed $M_J$ are 
non-diagonal in the $2\times 2$  sector with $J', J = N\pm \thalf$ and we need,
\begin{equation}
\la v\,N\,J'\,M_J|\,Y_{20}(\widehat{{\bs P}})\,|v\,N\,J\,M_J\ra ~=~
\sqrt{\frac{5}{4\pi}}\,\sum_{M_S}\,C_{N M_N,\,\thalf M_S}^{J' M_J} \, C_{N M_N,\,\thalf M_S}^{J M_J}\, c_{N M_N}\ .
\label{d14}
\end{equation}
Finding the eigenvalues in exactly the same way as described for the electron (see Appendix \ref{appA}) 
therefore gives
\begin{equation}
\D E_{\rm SME \,\pm}^n\big|_{n=0,2} ~=~ -\frac{2}{\sqrt{4\pi}}\,\, \VV_{000}^{\,p\,\NR} ~-~
2\,\la v\, N|\,P^2\,|v\, N\ra\, \Big(\frac{1}{\sqrt{4\pi}}\,\,\VV_{200}^{\,p\,\NR}
~+~ \sqrt{\frac{5}{4\pi}}\,\hat{c}_{N M_J}^\pm(B)\,\,\VV_{220}^{\,p\,\NR}\,\Big) \ .
\label{d15}
\end{equation}
with the same coefficients $\hat{c}_{N M_J}^\pm(B)$ introduced in (\ref{b26a}) and (\ref{b26b}).

The next step is to evaluate the expectation value $\la v\,N|\,P^2\,|v\,N\ra$ of the proton momentum 
in the spin-independent rovibrational states $|v\, N\ra$.  These states are constructed in a systematic
perturbation expansion about the simple harmonic oscillator (SHO) states found by expanding about the
minimum $R_0$ of the inter-nucleon potential $V_M(R)$. The details of this derivation are presented
in a self-contained way in Appendix \ref{appC}. The results are expressed in  a form directly matching
the expression in (\ref{aE}) for the full rovibrational energy spectrum, {\it i.e.}~an expansion
in $(v+\thalf)$ and $N(N+1)$ with coefficients parametrised by $\lambda \sim \sqrt{m_e/m_p}$.
We find,
\begin{align}
\la v\,N|\,P^2\,|v\,N\ra &=~ \half m_p \,\omega_0\,\bigg[(v+\thalf) \,+\, N(N+1)\,\l  \nonumber \\[2pt]
&~~~~~~~~~~~~~~~~~~~~~~~~~~~~~~~~
\,+\, \frac{3}{2}\Big(3 \,+\,\frac{R_0 V_M^{\prime\prime\prime}}{V_M^{\prime\prime}}\Big)
 \,(v+\thalf)\,N(N+1) \,\l^2 ~+\ldots \bigg] \nonumber \\[10pt]
&=~\half m_p \,\omega_0\,\bigg[(v+\thalf) \,+\, N(N+1)\,\l 
\,-\,3.405 \,(v+\thalf)\,N(N+1) \,\l^2 ~+\ldots \bigg] \ ,
\label{d16}
\end{align}
where $\omega_0$ is the frequency of the unperturbed SHO oscillations, given in terms 
of the Rydberg constant by $\omega_0 \sim R_H\sqrt{m_e/m_p}$. 

Together, (\ref{d16}) with (\ref{d14}) or (\ref{d15}) give the proton contributions at $O(P^2)$
to the components $\d_{\rm SME}^n$, $B_{\rm SME}^n$ and $\a_{\rm SME}^n$ in the rovibrational 
energy levels.\footnote{To match with the corresponding expression in terms of Lorentz tensor couplings
in the minimal SME, the dictionary is,
\begin{equation*}
\frac{1}{\sqrt{4\pi}}\,
\VV_{200}^{\,p\,\NR} ~= - \frac{1}{3 m_p^2}\,\tr E_{ij}^p \ , ~~~~~~~~~~~~~~~~
\sqrt{\frac{5}{4\pi}}\, \VV_{220}^{\,p\,\NR}~=~ \frac{1}{3m_p^2}\, \tr_Y E_{ij}^p \ ,
\end{equation*}
with \cite{Kostelecky:1999zh},
\begin{equation*}
E_{ij} ~=~ -m (c_{ij} + \thalf c_{00} \,\d_{ij}) \,+\, m^2 (3 a_{0ij} + a_{000}\, \d_{ij}) \ .
\end{equation*}
}
\vskip0.3cm
Now we have seen how to reformulate the derivation of the proton SME couplings to the
rovibrational energies at $O(P^2)$ entirely within the spherical tensor formalism, we are equipped 
to extend the analysis to the non-minimal SME, incorporating effects at $O(P^4)$ and
the associated non-minimal couplings.  At $O(P^4)$, the SME Hamiltonian is
\begin{equation}
H_{\rm SME}^p\big|_{n=4}  ~=~ - p_1^4 \,\Big( \frac{1}{\sqrt{4\pi}}\, \VV_{400}^{\,p\,\NR}
\,+\, Y_{2m}(\hat{\bs p}_1)\,\VV_{42m}^{\,p\,\NR} \,+\,
Y_{4m}(\hat{\bs p}_1)\,\VV_{44m}^{\,p\,\NR}\,\Big)  ~-~ ( \,{\bs p}_1 \,\rta\, {\bs p}_2\,) \ .
\label{d17}
\end{equation}

At this point we would hope to make the change of variables from ${\bs p}_1$, $\,{\bs p}_2$ 
to ${\bs P}$, $\,{\bs p}$ with identities analogous to (\ref{d6}) above.  However, even for the simplest
case, the coefficient of $\VV_{400}^{\,p\,\NR}$, we encounter a difficulty.  From the definitions (\ref{d2})
we immediately find,
\begin{equation}
p_1^4 \,+\, p_2^4 ~=~ 2 P^4 \,+\, P^2 p^2 \,+\, ({\bs P}\cdot{\bs p})^2 \,+\, \frac{1}{8} p^4 \ ,
\label{d18}
\end{equation}
so the mixed terms involving both ${\bs P}$ and ${\bs p}$ do {\it not} cancel. 
This spoils the clean separation of the total Schr\"odinger equation into electron and proton parts,
which underpins the simple Born-Oppenheimer split. We can try different strategies to evaluate
these mixed terms, but no sufficently simple results emerge for our present purpose. 
Systematic improvements on the basic Born-Oppenheimer have been the subject of extensive
theoretical work in atomic physics; for a review, see {\it e.g.}~\cite{Carrington:1989}.
We therefore neglect these terms
in what follows, and calculate the rovibrational energies in the approximation,\footnote{To estimate
the size of the neglected terms, note that $\la p^2\ra \sim R_H m_e$ while $\la P^2\ra \sim R_H \sqrt{m_e m_p}$,
so they will give corrections suppressed by $O(\sqrt{m_e/m_p}) \sim \l$ to the leading terms
quoted below separately for each coefficient $x_{\rm SME}^n$, $\,B_{\rm SME}^n$, $\,\a_{\rm SME}^n$, \ldots
in the expansion of the rovibrational energy.}
\begin{equation}
p_1^4\, Y_{jm}(\hat{\bs p}_1) \,+\, p_2^4 Y_{jm}(\hat{\bs p}) ~=~ 2\, P^4\, Y_{jm}(\widehat{\bs P}) 
~+~ \ldots 
\label{d19}
\end{equation}

We therefore take,
\begin{equation}
H_{\rm SME}^p\big|_{n=4} ~=~ -2 \,P^4\, \Big(\frac{1}{\sqrt{4\pi}}\, \VV_{400}^{\,p\,\NR}
\,+\, Y_{2m}(\widehat{\bs P})\,\VV_{42m}^{\,p\,\NR} \,+\,
Y_{4m}(\widehat{\bs P})\,\VV_{44m}^{\,p\,\NR}\,\Big) ~+~\ldots
\label{d20}
\end{equation} 
The corresponding energies are found as usual by taking expectation values in either the Zeeman
or hyperfine states for strong or weak magnetic fields respectively. Again, the expectation values
of the spherical harmonics yield coefficients already evaluated for the electron couplings, so we can 
immediately write the final results.
For strong magnetic fields,
\begin{align}
\D  E_{\rm SME\,\pm}^n \big|_{n=4} ~=~ &- 2\, \la v\,N|\,P^4\,|v\,N\ra   \nonumber \\[8pt]
&~~~\times~\Big(
\frac{1}{\sqrt{4\pi}}\,\VV_{400}^{\,p\,\NR}  \,+\, \sqrt{\frac{5}{4\pi}}\, c_{N M_J\mp\thalf} \, \VV_{420}^{\,p\,\NR} 
\,+\,\frac{3}{\sqrt{4\pi}}\, c_{NM_J\mp\thalf}^{(4)}\, \VV_{440}^{\,p\,\NR}\, \Big)\ ,
\label{d21}
\end{align}
while for weak fields,
\begin{align}
\D  E_{\rm SME\,\pm}^n \big|_{n=4} ~=~ &- 2\, \la v\,N|\,P^4\,|v\,N\ra   \nonumber \\[8pt]
&~~~\times \Big(
\frac{1}{\sqrt{4\pi}}\,\VV_{400}^{\,p\,\NR}  \,+\, \sqrt{\frac{5}{4\pi}}\, \hat{c}_{N M_J}^\pm \, \VV_{420}^{\,p\,\NR} 
\,+\, \frac{3}{\sqrt{4\pi}}\, \hat{c}_{NM_J}^{(4)\,\pm}\, \VV_{440}^{\,p\,\NR}\, \Big)\ ,
\label{d22}
\end{align}
with the coefficients quoted in (\ref{b20}) and (\ref{b27a}), (\ref{b27b}).

The momentum expectation value is evaluated in Appendix \ref{appC}, where we show
\begin{align}
\la v\,N|\,P^4\,|v\,N\ra ~=~  \frac{1}{4} \,m_p^2\,\omega_0^2\,\bigg[\, &\frac{3}{2} \,\big[(v+\thalf)^2 + \tfrac{1}{4}\big]
\,-\, \frac{219}{16}\,  N(N+1)\l^2  \nonumber \\[6pt]
&~~~~~~~~~~~~~~~~~~~~~~ + 2\,(v+\thalf)\,N(N+1)\,\l ~+~ \ldots\big]  \ ,~~~~
\label{d23}
\end{align}
where the $\dots$ denote terms of lower order in $\l$ which should be dropped for consistency given the
initial approximation in (\ref{d19}).
Notice the different orders in $\l$ of these terms compared with their equivalents in the expectation
value $\la v\,N|\,P^2\,|v\,N\ra$ in (\ref{d16}).

Together, (\ref{d23}) with (\ref{d21}) or (\ref{d22} give the proton contributions to the components
$x_{\rm SME}^n$, $\,B_{\rm SME}^n$ and $\a_{\rm SME}^n$ in the rovibrational energies at $O(P^4)$. 
Notice there is no contribution to $\d_{\rm SME}^n$ in this case.

These $O(P^4)$ contributions to the rovibrational energy coefficients provide sensitivity 
to the non-minimal SME couplings $\VV_{400}^{\,p\,\NR}$, $\,\VV_{420}^{\,p\,\NR}$ and $\VV_{440}^{\,p\,\NR}$.
Individual constraints on these, as well as $\VV_{200}^{\,p\,\NR}$ and $\VV_{220}^{\,p\,\NR}$,
may be achieved by measurements of spectral transitions corresponding to
states with different rovibrational quantum numbers $(v,N)$ through the leading terms in 
(\ref{d16}) and (\ref{d23}) and, crucially, through discriminating the different spin components
$\VV_{njm}^{\,p\,\NR}$ by exploiting the quantum number dependence of the $c$ and $\hat{c}$ coefficients.

\section{Frame transformations and transition frequency variations}\label{sect 5}

In the SME, Lorentz symmetry is broken because the vector and tensor couplings (which may be usefully 
thought of as VEVs of fields in some more fundamental theory) take on fixed, non-zero values.
Unlike the familiar case of QFTs with scalar couplings, this means that in the SME the results of
physical measurements depend on the frame of reference of the experiment.

This has two immediate consequences. First, in comparing the results of experiments at different 
locations and orientations, we need to make an appropriate Lorentz transformation (rotation or boost)
to change the frame of reference of the individual experiments. Second, measurements in the same
experiment at a fixed location on Earth will experience daily (sidereal) and annual variations
as the frame of reference of the experiment changes as the Earth rotates and orbits the Sun.

To compare the results of different experiments, or the same experiment performed at different times,
it is convenient to refer the couplings to a standard frame \cite{Kostelecky:2008ts}. 
It has become established practice to use a solar-centred frame (\textsf{SUN}), defined by an 
orthornormal set of basis vectors ${\bs e}_I$ with the $Z$-axis aligned with the Earth's rotation axis 
and the $X$-axis with the vernal equinox.
This implies that ${\bs e}_X,\,{\bs e}_Y$ lie in the plane of the Earth's equator,
making an angle of $\eta = 23^\circ$ with the orbital plane. Where necessary, the origin of the time 
coordinate T is chosen as the vernal equinox in 2000. 

In practice it is convenient to introduce one further frame,  a standard laboratory frame (\textsf{LAB}).
The \textsf{LAB} frame is defined locally with basis vectors ${\bs e}_i$ with the $z$-axis pointing vertically 
to the zenith at the location of the experiment and the $x$-axis pointing south. 
The relation of the ${\bs e}_z$ and ${\bs e}_Z$ axes in the \textsf{LAB} and \textsf{SUN} frames is
given by the colatitude $\chi$ of the laboratory.

\subsection{Rotations and daily (sidereal) variations}\label{sect 5.1}

Since our results for the energy levels are given in terms of couplings in \textsf{EXP} frame coordinates,
we need to convert to the standard \textsf{SUN} frame in two stages, first re-expressing in terms
of the \textsf{LAB} frame then a further rotation from \textsf{LAB} to \textsf{SUN}. 
We illustrate this transformation for a generic spin-independent coupling $\VV_{njm}$, dropping all  
other labels since the rotation is entirely determined by the $SO(3)$ indices $(j,m)$. The results for
the spin-dependent couplings $\TT_{njm}$, of all types, are identical.

To transform from \textsf{EXP} to \textsf{LAB}, we need the angle $\phi$ of the magnetic field measured 
anticlockwise from due south, and its angle $\theta$ measured from the local vertical. 
The couplings transform with the Wigner matrices as (following our notation so far, couplings with
no superscript are in the \textsf{EXP} frame), 
\begin{equation}
\VV_{njm'} ~=~ \VV_{njm}^{\LAB} \, e^{i m \phi}  \, d_{m m'}^{\,j}(\theta)  \ .
\label{e1}
\end{equation}
Combining with a further rotation to the \SUN frame we can show (non-trivially, since in both cases
the angular momentum operators defining the Wigner matrices are to be taken in the initial frame
\cite{VarshalovichBook}), that
\begin{equation}
\VV_{njm'}~=~ \VV_{njm}^{\SUN}\, e^{i m \omega_{\oplus} T_{\oplus}} \, d_{m m^{\prime\prime}}^{\,j}(\chi)\,
e^{i m^{\prime\prime} \phi} \, d_{m^{\prime\prime} m'}^{\,j} (\theta) \ ,
\label{e2}
\end{equation}
where $\omega_{\oplus}$ is the Earth's angular rotation frequency corresponding to a sidereal day
and $T_{\oplus}$ being local sidereal time in the \LAB frame. 

Next, we use the Wigner matrix addition theorem\footnote{
The standard form for the addition theorem for the little Wigner matrices $d_{m m'}^{\,j}$ is 
\cite{VarshalovichBook}
\begin{equation*}
d_{m m^{\prime\prime}}^{\,j}(\b_1) \, e^{- i m^{\prime\prime}\varphi}\, d_{m^{\prime\prime} m'}^{\,j}(\b_2) 
~=~ e^{-i m\a}\, d_{m m'}^{\,j}(\b) \, e^{-i m' \gamma} \ ,
\end{equation*}
with
\begin{align*}
\cot\a ~&=~ \cos\b_1\, \cot\varphi \, +\,  \sin\b_1\,\cot\b_2 \,(\sin\varphi)^{-1} \ , \\
\cos\b ~&=~  \cos\b_1\,\cos\b_2\,- \,\sin\b_1\,\sin\b_2\,\cos\varphi \ , \\
\cot\gamma ~&=~ \cos\b_2\,\cot\varphi\,+\,sin\b_2\,\cot\b_1\,(\sin\varphi)^{-1} \ . 
\end{align*}
Also note the useful relation $d_{m m'}^{\,j}(-\theta) \,=\, d_{m' m}^{\,j}(\theta)$.} to show,
\begin{equation}
\VV_{njm'} ~=~ \VV_{njm}^{\SUN} \, e^{im(\omega_\oplus T_\oplus + \a)}  \,  d_{m m'}^{\,j}(\vartheta) 
\, e^{i m' \gamma}  \ ,
\label{e3}
\end{equation}
where here,
\begin{align}
\cos\vartheta ~&=~  \cos\chi\, \cos\theta  \,-\, \sin\chi\, \sin\theta\, \cos\phi  \ , 
\label{e4} \\[5pt]
\cot\a ~&=~ \cos\chi\, \cot\phi \,+\,  \sin\chi \,\cot\theta \, (\sin\phi)^{-1} \ ,
\label{e5}
\end{align}
with an identical formula for $\cot\gamma$ with $\chi,\, \theta$ interchanged.

Next, recalling that only couplings with $m=0$ in the \EXP frame appear in the energy levels, 
we only need
\begin{align}
\VV_{nj0}^{w\,\NR} ~&=~  \VV_{njm}^{w\,\NR\,\SUN} \, d_{m 0}^{\,j}(\vartheta) \, e^{i m (\omega_\oplus T_\oplus + \a)} \ ,
\nonumber \\[6pt]
\TT_{nj0}^{w\,\NR} ~&=~ \TT_{njm}^{w\,\NR\,\SUN} \, d_{m 0}^{\,j}(\vartheta) \, e^{i m (\omega_\oplus T_\oplus + \a)} \ ,
\label{e6}
\end{align}
where we restore the full notation, with $w = e,p$.
These expressions are to be inserted into the formulae for the inter-nucleon potential and energy levels
to give our final results in the standard \SUN frame.

We see immediately that rovibrational transition frequencies measured at different times during the
Earth's daily rotation will show a time variation. Moreover, this depends on {\it all} the azimuthal
components of the couplings, not just $m=0$.  This is quite natural in a Lorentz violating theory --
measurements taken with a different orientation of the experiment will be sensitive to different 
components of a background Lorentz vector or tensor, such as the effective SME couplings. 

In principal, the dependence of these variations on the couplings with different values of $m$ can be 
isolated since they vary according to different harmonics of the rotation frequency, arising
from the $\exp\big(im(\omega_\oplus T_\oplus + \a)\big)$ factor.

Notice also that the orientation of the magnetic field also introduces a constant, experiment-specific, phase 
into the variation. This will not be important in analysing time variations in a single experiment, since only
time differences will matter, but could be relevant in comparing results from experiments with different
magnetic field orientations or in different laboratories, since according to (\ref{e5}) the phase $\a$
depends on $\phi, \theta$ and $\chi$.

We can make (\ref{e6}) more explicitly tailored to experimental measurements by rewriting entirely 
in terms of real quantities. Using the relations $\VV_{nj (-m)}\,=\, (-1)^m\,\VV_{njm}^*$ 
and $d_{(-m) 0}^{\,j} \,=\,(-1)^m\,d_{m0}^{\,j}$, (\ref{e6}) becomes,
\begin{align}
&\VV_{nj0}^{w\,\NR}~=~ d_{00}^{\,j}(\vartheta)\, \VV_{nj0}^{w\,\NR\,\SUN} \nonumber \\[6pt]
&+
\sum_{|m|=1}^j\, d_{|m|\,0}^{\,j}(\vartheta) \, \Big(\cos\big(m(\omega_\oplus T_\oplus + \a)\big)\, \Re \VV_{nj|m|}^{w\,\NR\,\SUN}
\,-\, \sin\big(m(\omega_\oplus T_\oplus + \a)\big)\, \Im \VV_{nj|m|}^{w\,\NR\,\SUN} \Big) \ .
\label{e7}
\end{align}

Finally, to illustrate how these time variations enter our final results, we quote here the explicit example
of the proton contribution to the energy levels, up to $O(P^2)$ and for weak fields, written out fully
in terms of \SUN frame couplings (see (\ref{d15})):\footnote{The corresponding expressions for many of the 
leading terms for a particular experimentally important transition frequency are tabulated
in \cite{Vargas:2025efi} and may be derived directly from the more general results given here.}
\begin{align}
&\D E_{{\rm SME}\,\pm}^n\big|_{n=0,2} ~=~ -\frac{2}{\sqrt{4\pi}}\,\, \VV_{000}^{\,p\,\NR\,\SUN} \nonumber \\
~&~~~~  -~2\,\la v\, N|\,P^2\,|v\, N\ra\, \bigg[\frac{1}{\sqrt{4\pi}}\,\,\VV_{200}^{\,p\,\NR\,\SUN}  
~ +~ \sqrt{\frac{5}{4\pi}}\,\hat{c}_{N M_J}^\pm(B)\,\,
\bigg(-\frac{1}{2}(1 - 3 \cos^2\vartheta) \VV_{220}^{\,p\,\NR} \nonumber \\
&~~~~ -~\frac{1}{2} \sqrt{\frac{3}{2}}\,\sin 2\vartheta \,\Big(
\cos(\omega_\oplus T_\oplus + \a) \,\Re \VV_{221}^{\,p\,\NR\,\SUN} 
\,-\, \sin(\omega_\oplus T_\oplus + \a) \,\Im \VV_{221}^{\,p\,\NR\,\SUN} \Big) \nonumber \\
&~~~~ +~\frac{1}{2} \sqrt{\frac{3}{2}}\, \sin^2\vartheta\,\Big(
\cos 2(\omega_\oplus T_\oplus + \a) \,\Re \VV_{222}^{\,p\,\NR\,\SUN} 
\,-\, \sin 2(\omega_\oplus T_\oplus + \a) \,\Im \VV_{222}^{\,p\,\NR\,\SUN} \Big)
\,\bigg) \bigg] \ .
\label{e8}
\end{align}

All the other expressions for energy levels in the paper can be expanded out in the same way showing the
dependence on all the azimuthal components of the spherical tensor couplings at fixed rank $j$.
As we now show, consideration of the effects of Lorentz boosts also exposes a time-dependence on couplings
of different rank.

\subsection{Lorentz boosts}\label{sect5.2}

The energy levels will also experience variations due to Lorentz boosts arising from the motion 
of the laboratory in the \SUN frame. These are of two types, daily (sidereal) variations determined
by the Earth's rotational velocity ${\bs v}_E$ and annual variations governed by the orbital 
velocity ${\bs v}_\oplus$.
Here \cite{Kostelecky:2015nma}, taking $c=1$, 
\begin{equation}
{\bs v}_E ~=~ v_E \,\big(- \sin\omega_\oplus T_\oplus\,\,{\bs e}_X \,+\, 
\cos\omega_\oplus T_\oplus\,\,{\bs e}_Y  \big)\ ,
\label{e9}
\end{equation}
while
\begin{equation}
{\bs v}_\oplus ~=~ v_\oplus\, \big( \sin\Omega_\oplus T\,\,{\bs e}_X \,-\, \cos\Omega_\oplus T\,
(\cos\eta \,\,{\bs e}_Y \,+\, \sin\eta\, {\bs e}_Z )\big) \ .
\label{e10}
\end{equation}
These velocities are small, with $v_E = O(10^{-6})$ and $v_\oplus = O(10^{-4})$,  so we may consider
the corresponding Lorentz boosts to $O(v)$ only ({\it i.e.} set the relativistic $\gamma$ factors to 1).

Now, while the spherical tensor formalism used here for the SME couplings is extremely convenient
in considering the effects of rotations, it is not well-suited to analyse Lorentz boosts. In particular,
spherical tensors $\VV_{njm}$ characterised by the same $(j,m)$ labels may have quite different
transformations under Lorentz boosts, since they only distinguish the rotational properties of their
constituent 4-vectors and tensors in the fundamental theory.
For example, even if we consider the simplest case of the $\VV_{njm}^{\NR} = c_{njm}^\NR - a_{njm}^\NR$
couplings, we find different transformation properties under boosts for the $c_{njm}^\NR$
and $a_{njm}^\NR$ separately. 

We therefore need to re-express the spherical tensor couplings in their original Lorentz tensor form 
to study the effects of Lorentz boosts.
We illustrate this first by evaluating the Lorentz boost transformations of the $c_{njm}^{\NR\,\SUN}$ couplings,
which are related to the original $c_{\m\n}$ tensor couplings.  This demonstrates most of the
general features in considering boosts with the spherical tensors.

First we need a more complete dictionary of the equivalence of the spherical and Lorentz tensor 
couplings than given in Paper 2. In the \SUN frame, we find \cite{Kostelecky:2013rta}
\begin{align}
\frac{1}{\sqrt{4\pi}}\, c_{000}^{\NR\,\SUN} ~&=~ m\, c_{TT}    \ , \nonumber \\[6pt]
\sqrt{\frac{3}{4\pi}}\, c_{11\,m=0,1,-1}^{\NR\,\SUN} ~&=  ~2\,c_{TZ}, ~~ -\sqrt{2}\, (c_{TX} -i c_{TY}), ~~
\sqrt{2}\, (c_{TX} +i c_{TY}) \ , \nonumber \\[6pt]
\frac{1}{\sqrt{4\pi}}\, c_{200}^{\NR\,\SUN} ~&=~ \frac{1}{3m}\,\big(c_{KK} \,+\, \frac{3}{2} c_{TT}\big) \ ,\nonumber \\[6pt]
\sqrt{\frac{5}{4\pi}}\, c_{220}^{\NR\,\SUN} ~&=~ -\frac{1}{3m}\,\big(c_{XX} \,+\, c_{YY}\,-\, 2c_{ZZ} \big) \ ,
\label{e11}
\end{align}
while more generally,
\begin{equation}
c_{22m}^{\NR\,\SUN} ~=~  \frac{1}{m}\, C_{JK}^{\,m}\, c_{JK} \ ,
~~~~~~~~~~
\label{e12}
\end{equation}
where $C_{JK}^{\,m}$ are the expansion coefficients for the product of two vectors in spherical harmonics
given in \cite{Yoder:2012ks}.\footnote{The $C^m_{JK}$ are defined from the expansion
$p_J p_K = p^2\big(\tfrac{1}{3} \d_{JK} + C_{JK}^m Y_{2m}\big)$.  
Explicitly, 
\begin{equation*}
C_{JK}^{\,0} \,=\, \sqrt{\frac{2\pi}{15}}\sqrt{\frac{2}{3}}\, 
\begin{pmatrix}1 &~0&~0\\0&~1&~0\\0&~0&-2\end{pmatrix} \ , ~~~~~~~~~~
C_{JK}^{\,1}\,=\,  \sqrt{\frac{2\pi}{15}}
\begin{pmatrix}~0 &~0&-1\\~0&~0&~i\\-1&~i&~0\end{pmatrix} \ , ~~~~~~~~~~
C_{JK}^2 \,=\, \sqrt{\frac{2\pi}{15}}
\begin{pmatrix}~1 &-i&~0\\-i&-1&~0\\0&~0&~0\end{pmatrix} \ .
\end{equation*}
}

Under a Lorentz boost the transformations of the spherical tensors entering the energy level
formulae are,
\begin{align}
\frac{1}{\sqrt{4\pi}}\,\, \d \, c_{000}^{\NR\,\SUN} ~&=~ 2m\,v^J\,c_{TJ} \ , ~~~~~~~~~~~~~~~~~~~~~~
\label{e13a}\\[6pt]
\frac{1}{\sqrt{4\pi}}\,\, \d \, c_{200}^{\NR\,\SUN} ~&=~  \frac{5}{3m}\, v^J\, c_{TJ} \ ,~~~~~~~~~~~~~~~~~~
\label{e13b}
\end{align}
and 
\begin{equation}
\sqrt{\frac{5}{4\pi}}\,\, \d \, c_{220}^{\NR\,\SUN} ~=~  -\frac{2}{3m}\, \big(v^J\, c_{TJ}  \,-\, 3\,v^Z\, c_{TZ}\big) \ .
\label{e14}
\end{equation}

Now, since in this paper we are expressing everything in the spherical tensor formalism, the next step is
to rewrite these in terms of the $c_{njm}^{\NR\,\SUN}$ couplings.
First, writing the velocity in the \SUN frame in its spherical vector form with components
$v_0 = v_Z, \, v_{m=\pm1} = \mp\tfrac{1}{\sqrt{2}}(v_X \pm i v_Y)$, 
we find using (\ref{e11}),
\begin{align}
\d\, c_{000}^{\NR\,\SUN} ~&=~ -\sqrt{3}\,m\,v_m\,c_{11m}^{\NR\,\SUN} \ , 
\label{e15a} \\[6pt]
\d\, c_{200}^{\NR\,\SUN} ~&=~ -\frac{5}{2\sqrt{3}}\,\frac{1}{m}\, v_m\, c_{11m}^{\NR\,\SUN}  \ ,
\label{e15b}
\end{align}
and 
\begin{equation}
\d\, c_{220}^{\NR\,\SUN} ~=~ \frac{1}{\sqrt{15}} \,\frac{1}{m}\,\big(
v_m\, c_{11m}^{\NR\,\SUN} \,-\, 3 v_0\, c_{110}^{\NR\,\SUN}\big) \ .
\label{e16}
\end{equation}

For the velocities in  (\ref{e9}), (\ref{e10}) we have,
\begin{align}
v_{E\,0} ~&=~ 0 \, \nonumber\\[2pt]
v_{E\, m} ~&=~ -\frac{i}{\sqrt{2}}\, v_E \,  e^{i m\, \omega_\oplus T_\oplus} ,    ~~~~~~~(m=\pm1) \ ,
\label{e17} 
\end{align}
and 
\begin{align}
v_{\oplus\,0} ~&=~ - \sin\eta \, \cos\Omega_\oplus T \ , \nonumber \\[4pt]
v_{\oplus\,m} ~&=~ \frac{i}{2\sqrt{2}} \Big[(1 + \cos\eta) e^{im\,\Omega_\oplus T}
\,-\,(1 -\cos\eta) e^{-i m\,\Omega_\oplus T} \Big] , ~~~~~~~(m=\pm1) \ .
\label{e18}
\end{align}
It follows that for sidereal variations, the non-isotropic term in (\ref{e16}) does not contribute.

We immediately see from (\ref{e15b}), (\ref{e16}) that the Lorentz boost has introduced a sensitivity 
to a new set of spherical tensor couplings, in this case $c_{11m}^{\NR\,\SUN}$, which did not appear 
in the original energy level formulae.
The numerical factors are largely specific to the $c_{njm}^{\NR\,\SUN}$ couplings, but we can understand
the qualitative features that determine which couplings arise.
First, they must satisfy the angular momentum addition rules for the combination of the couplings
with the $(1,m)$ components of the vector $v_m$. 
Then, the $n$ label must change by $\pm 1$ to maintain the matching with the power of $|{\bs p}|^n$
in the non-relativistic Hamiltonian defining the couplings. 
Finally, the range of allowed $j$ is further constrained by the restrictions $j = n, n-2,\ldots \ge 0$
on the $c_{njm}^{\NR}$ couplings \cite{Kostelecky:2013rta}.
Taken together, these constraints explain the occurrence of the $c_{11m}^{\NR\,\SUN}$ coupling in the
boosts of $c_{200}^{\NR\,\SUN}$ and $c_{220}^{\NR\,\SUN}$. The non-isotropy of $c_{220}^{\NR\,\SUN}$
is apparent in the independent presence of $c_{110}^{\NR\,\SUN}$ in (\ref{e16}).

Once again, as in (\ref{e8}) for $\D E_{\rm SME\,\pm}^n$, these transformations should be inserted into
the expressions for the energy levels to see the effect of the boost-induced sidereal or annual
variations, which may in principle be used to isolate the dependence on the extra couplings
$c_{11m}^{\NR\,\SUN}$.

The obvious question now is whether the same results apply also to the corresponding \textsf{CPT}
odd couplings $a_{njm}^{\NR\,\SUN}$ which so far we have treated together in the coupling
$\VV_{njm}^{\NR\,\SUN}$.  However, here the answer is no. These couplings arise from a different
tensor $a_{\m\n\l}$ in the fundamental Lagrangian. This is of rank 3 compared to the rank 2 $c_{\m\n}$
and the couplings therefore have quite different Lorentz transformations.

In this case, comparing the non-relativistic Hamiltonian with the fundamental Lagrangian, 
and focusing just on the dimension 5 $a_{\m\n\l}$ couplings ({\it i.e.}~ignoring the dimension 3 $a_\m$),
we can show:
\begin{align}
\frac{1}{\sqrt{4\pi}}\, a_{000}^{\NR\,\SUN} ~&=~ m^2\, a_{TTT}    \ , \nonumber \\[6pt]
\sqrt{\frac{3}{4\pi}}\, a_{11\,m=0,1,-1}^{\NR\,\SUN} ~&=  ~-3m\,a_{TTZ},
 ~~ 3m\,a_{TT-}, ~~-3m\, a_{TT+}\ , \nonumber \\[6pt]
\frac{1}{\sqrt{4\pi}}\, a_{200}^{\NR\,\SUN} ~&=~ a_{TTT} \,+\, a_{TKK} \ ,\nonumber \\[6pt]
\sqrt{\frac{5}{4\pi}}\, a_{220}^{\NR\,\SUN} ~&=~ -\big(a_{TXX} \,+\, a_{TYY}\,-\, 2a_{TZZ} \big) \ ,
\label{e19}
\end{align}
while more generally,
\begin{equation}
a_{22m}^{\NR\,\SUN} ~=~  3 \, C_{JK}^{\,m}\, a_{TJK}\ ,
~~~~~~~~~~
\label{e20}
\end{equation}
with $C_{JK}^m$ as above. Here, we define $a_{TT\pm} =\tfrac{1}{\sqrt{2}}\,( a_{TTX} \pm i a_{TTY})$.

The Lorentz boosts give,
\begin{align}
\frac{1}{\sqrt{4\pi}}\,\,\d\, a_{000}^{\NR\,\SUN} ~&=~ 3 m^2\,v^J\,a_{TTJ}\ , 
\label{e21a} \\[6pt]
\frac{1}{\sqrt{4\pi}}\,\,\d\, a_{200}^{\NR\,\SUN} ~&=~ 5 \,v^J\,a_{TTJ} \, + \,  v^J\, a_{JKK} \ ,
\label{e21b}
\end{align}
and 
\begin{equation}
\sqrt{\frac{5}{4\pi}}\,\,\d\, a_{220}^{\NR\,\SUN} ~=~ -2\,v^J\,a_{TTJ} \,-\, v^J\,a_{JKK} \,+\, 3\,v^J\,a_{JZZ}
\,+6\,v^Z\,a_{TTZ} \ .
\label{e22}
\end{equation}

Re-expressing in spherical tensor form we immediately find
\begin{equation}
\d\,a_{000}^{\NR\,\SUN} ~=~ \sqrt{3}\,m\,v_m\, a_{11m}^{\NR\,\SUN} \ .
\label{e23}
\end{equation}
Compared to $\d\,c_{200}^{\NR\,\SUN}$, the boost for $a_{200}^{\NR\,\SUN}$ has an extra type
of term proportional to the purely space components $a_{IJK}$ of the fundamental tensor. 
In general, writing these in spherical tensor form involves the couplings $a_{3jm}^{\NR\,\SUN}$ with
$j=1,3$.  This case is simplified, however, since we only require the couplings with contracted indices
and we can straightforwardly check,
\begin{equation}
\sqrt{\frac{3}{4\pi}} \,\big(a_{31m}^{\NR\,\SUN}  \,-\, \frac{1}{2m^2}\, a_{11m}^{\NR\,\SUN}\big)_{m=0,1,-1}
~=~ -\frac{3}{5m}\, a_{KKZ}, ~~~~~~ \frac{3}{5m}\, a_{KK-}, ~~~~~~-\frac{3}{5m}\, a_{KK+} \ .
\label{e24}
\end{equation}
Then, collecting terms,
\begin{equation}
\d\,a_{200}^{\NR\,\SUN} ~=~ \frac{5}{\sqrt{3}} \,m\,v_m\, \big(a_{31m}^{\NR\,\SUN} \,+\, 
\frac{1}{2m^2}\,a_{11m}^{\NR\,\SUN} \big) \ .
\label{e25}
\end{equation}
We therefore find a contribution in which $n$ has {\it increased} by 1. This is due
to the rank of the underlying fundamental Lorentz coupling. The allowed $(j,m)$ labels on the
couplings are constrained by the angular momentum addition rules as before.

Turning finally to $\d\,a_{220}^{\NR\,\SUN}$, we also encounter a non-isotropic tensor coupling $a_{JZZ}$.
Inspection of the matching conditions from the Hamiltonians shows that this also involves
the $j=3$ couplings $a_{33m}^{\NR\,\SUN}$.  After some further calculation we find
\begin{equation}
\sqrt{\frac{7}{4\pi}}\, a_{330}^{\NR\,\SUN} ~+~ \sqrt{\frac{3}{4\pi}} \,\Big(a_{310}^{\NR\,\SUN} \,-\, 
\frac{1}{2m^2}\,a_{110}^{\NR\,\SUN}  \Big)  ~=~ -\frac{1}{m}\, a_{ZZZ} \ ,
\label{e26}
\end{equation}
while
\begin{equation}
\sqrt{3}\,\sqrt{\frac{7}{4\pi}}\, a_{33m}^{\NR\,\SUN} ~+~ \sqrt{\frac{3}{4\pi}} \,\Big(a_{31m}^{\NR\,\SUN} \,-\, 
\frac{1}{2m^2}\,a_{11m}^{\NR\,\SUN}  \Big)  ~=~ \pm\frac{3}{m}\, a_{ZZ\mp} ,   ~~~~~~(m=\pm1) \ .
\label{e27}
\end{equation}

Then, from (\ref{e22}) and collecting terms, we can write the Lorentz boost for the coupling 
$a_{220}^{\NR\,\SUN}$ as,
\begin{align}
\d\, a_{220}^{\NR\,\SUN}  ~&=~~
m\,v_m\,\Big( \sqrt{\frac{21}{5}}\,\,a_{33m}^{\NR\,\SUN}  ~-~ 
\frac{2}{3}\,\sqrt{\frac{3}{5}}\,\,\big(a_{31m}^{\NR\,\SUN}  \,+\, \frac{1}{2m^2}\, a_{11m}^{\NR\,\SUN}\big)\Big)
\nonumber \\
&~~-3 m\,v_0\,\Big( \big(\frac{1}{\sqrt{3}}-1\big) \sqrt{\frac{7}{5}}\,\, a_{330}^{\NR\,\SUN}  ~-~ 
\frac{2}{3}\sqrt{\frac{3}{5}}\,\,\big(a_{310}^{\NR\,\SUN}  \,+\, \frac{1}{2m^2}\, a_{110}^{\NR\,\SUN}\big)\Big) \ .
\label{e28}
\end{align}
This is to be compared with the corresponding result (\ref{e16}) for the Lorentz boost of $c_{220}^{\NR\,\SUN}$,
the key difference being the appearance of couplings with $n=3$.

Evidently, we can continue to evaluate the Lorentz boosts for all the spherical tensor couplings 
appearing in the relations for the energy levels.  Each must be evaluated separately since the
underlying fundamental Lorentz tensors will be different in each case.
The general principles and all the rules governing which couplings can appear are the same as 
demonstrated above for the $c_{njm}^{\NR\,\SUN}$ and $a_{njm}^{\NR\,\SUN}$ couplings
and we will not quote all of these explicitly here.

The key implication for a future programme of high-precision spectroscopy is that the SME predicts
sidereal, and annual, variations of transition frequencies which, although suppressed by
$O(v_E)$ or $O(v_\oplus)$, depend on {\it extra} spherical tensor couplings with $n$ and $j$ differing 
by 1 compared to those which would appear in time-averaged measurements.

\vskip1cm
\section{Outlook}\label{sect 6}

In this paper, extending the work presented in Papers 1 and 2, we have provided a comprehensive account of
the effect of the Lorentz and \textsf{CPT} violating interactions in the SME on the rovibrational spectrum 
of the molecular ions \Hmol and \Hbarmol.
The description has been entirely in terms of the spherical tensor representation of the couplings,
which is the most natural framework to include the non-renormalisable operators in the non-minimal SME.

We have given our results as perturbations on the full hyperfine-Zeeman spectrum, for both large and small
background magnetic fields. This makes them directly applicable both to current high-precision 
spectroscopy of \Hmol using low-field RF traps \cite{SAS2024} and to future spectroscopy of its antimatter
counterpart \Hbarmol, which will be performed in a high magnetic field on molecular ions confined in a 
Penning trap \cite{SchillerCERN2026}.

The effect of the Lorentz and \textsf{CPT} violating interactions involving the proton are calculated directly 
in perturbation theory. For the electron interactions, their direct effect is to modify the inter-nucleon potential.
Following Paper 1, we showed how a detailed analysis of the molecular dynamics, noting particularly
the lack of full spherical symmetry of the molecule even in its $1s\s_g$ electron ground state,
allows this to be translated into changes to the rovibrational energy levels. We also noted how in the
Born-Oppenheimer framework, the proton SME couplings also feed into the inter-nucleon potential,
and subsequently the rovibrational spectrum, via (\ref{d10}). 

In both cases, we quoted the changes to the rovibrational energy levels in complete generality, giving 
the dependence of the SME contributions on the quantum numbers $(v,N)$ and spin $M_J$
of the hyperfine-Zeeman levels at high and low magnetic field.

We then described the importance of looking for daily (sidereal) and annual variations of the transition
frequencies induced by rotations and Lorentz boosts in this Lorentz non-invariant theory. 
These variations are especially significant as they depend on further components of the SME couplings
characterised by different $(njm)$ labels, which would not be present in single or time-averaged 
measurements.

All this establishes the theoretical basis to interpret, and guide, experimental searches for Lorentz 
and \textsf{CPT} violation in hydrogen and antihydrogen molecular ion spectroscopy.

Rovibrational spectroscopy of \Hmol at $O(10^{-12})$ has already been carried out \cite{SAS2024}
with the possibility of eventually attaining precisions as high as $O(10^{-17})$. Meanwhile, the success
of the ALPHA collaboration in high-precision spectroscopy of atomic antihydrogen, especially the
$1S$\,-\,$2S$ transition \cite{Ahmadi:2018eca, Baker:2025ehs}, 
together with the recent achievement of BASE-STEP in transporting 
antiprotons away from the electromagnetically noisy environment of the CERN AD/ELENA facility,
has opened an increasingly clear path towards future high-precision spectroscopy of \Hbarmol.

An important element in achieving such precisions in molecular ion spectroscopy is to put together
combinations of specific rovibrational transitions chosen so as to cancel the systematic shifts due
to less precisely controllable effects such as the linear and quadratic Zeeman shifts, the electric 
quadrupole shift, and dependence on the hyperfine spin structure \cite{SchillerKorobov2018}. 
Each of these has a distinctive dependence on the $v, N$ and $M_J$ quantum numbers
\cite{Korobov2006, Karr2008, KKH20081}, 
ultimately determined, like the SME contributions, by combinations of Clebsch-Gordan coefficients.

This raises a serious question for the design of experimental protocols to search for Lorentz and 
\textsf{CPT} violation.  Inspection of the quantum number dependence of our results for the 
SME contributions, encoded in the $c_{NM_J}$, $e_{NM_J}$ and $f_{NM_J}$ coefficients, shows that in selecting 
combinations of transition frequencies that cancel the systematic shifts listed above, sensitivity to some
or all of the SME couplings could be lost. In the most unfavourable case, an extremely precise transition 
frequency combination may be entirely insensitive to CPT violation.

An important next step is therefore to use our results to analyse in detail possible specific combinations of
rovibrational transition frequencies which minimise sensitivity to precision-limiting systematic effects
while maintaining maximum sensitivity to potential sources of Lorentz and \textsf{CPT} violation. 
This analysis will be presented in forthcoming work.

\vskip0.3cm
\noindent {\large{\textbf{Acknowledgements}}}
\vskip0.3cm
I am grateful to Stefan Eriksson for many interesting discussions on the potential of
antihydrogen molecular ion spectroscopy, 
and to the Higgs Centre for Theoretical Physics at the University of Edinburgh for hospitality.

\newpage

\appendix{

\section{Hyperfine-Zeeman Hamiltonian and eigenvalues of $V_{\rm SME}^e(R)$}\label{appA}

The hyperfine-Zeeman Hamiltonian may be written as
\begin{equation} 
H_{HFS} \,+\,H_Z ~=~ c_e(v,N)\,{\bs N}\cdot{\bs S} \,+\, g_e \mu_B\, {\bs B}\cdot{\bs S}
\,-\, g_m(v,N) \mu_B\, {\bs B}\cdot{\bs N} \ ,
\label{appA1}
\end{equation} 
where the couplings $c_e(v,N)$ and $g_m(v,N)$ depend on the rovibrational state.
Its expectation value in the states $|v\,N\,J\,M_J\ra$ can be written for fixed $M_J$ (which remains
a good quantum number in the presence of the applied magnetic field ${\bs B}$)
as a $2 \times 2$ matrix with rows/columns corresponding to $J,\,J' = N\pm \thalf$ as follows,
\begin{equation} 
\la v\,N\,J'\,M_J|\,H_{HFS} \,+\,H_Z \,|v\,N\,J\,M_J\ra ~=~ 
\begin{pmatrix}\textsf{A} ~&\textsf{B} \\ \textsf{B} ~&\,\textsf{D} \end{pmatrix}\ ,
\label{appA2}
\end{equation} 
with
\begin{align}
A ~&=~ \frac{1}{2} N c_e(v,N) \,+\, \frac{1}{2N+1} \,\big(g_e - 2N g_m(v,N)\big)\,\mu_B B M_J \ ,
\nonumber \\[4pt]
D ~&=~ - \frac{1}{2} (N+1) c_e(v,N) \,-\, \frac{1}{2N+1}\,\big(g_e + 2(N+1) g_m(v,N)\bigr) \mu_B B M_J \ ,
\nonumber \\[4pt]
B ~&=~ - \frac{1}{2N+1} \sqrt{(N+\thalf)^2 - M_J^2}\,\,
 \big(g_e + g_m(v,N) \big) \m_B B \ .
\label{appA3}
\end{align}

With the matrix elements $V_{\rm SME}^e(R)$ in the same states written as
\begin{equation}
V_{\rm SME}^e(R) ~=~ \begin{pmatrix}\a ~&\b\\ \b^* ~&\,\d \end{pmatrix}\ ,
\label{appA4}
\end{equation}
the SME contributions to the combined eigenvalues are given by (Paper 2, footnote 9),
\begin{equation}
V_{\rm SME\,\pm}^e(R) ~=~ \thalf (\a + \d) ~\pm~ \thalf \, \frac{1}{\sqrt{(A-D)^2 + 4B^2}}
\,\big((A-D)(\a-\d) + 2B(\b+\c)\,\big) \ .
\label{appA5a}
\end{equation}
In the case considered here, the off-diagonal element \textsf{B} is proportional to the magnetic
field, so in the small-field regime where we can neglect terms of
$O(\mu_B B/c_e)^2$, we have the
simple forms:
\begin{equation}
V_{\rm SME\,+}^e(R) ~=~ \a \,+ \, 2\,\frac{\textsf{B}}{\textsf{A}-\textsf{D}}\, \Re \b \ , ~~~~~~~~~~
V_{\rm SME\,-}^e(R) ~=~ \d \,- \, 2\,\frac{\textsf{B}}{\textsf{A}-\textsf{D}}\, \Re \b  \ .
\label{appA5}
\end{equation}
From above, we find at $O(B)$,
\begin{equation}
\frac{\textsf{B}}{\textsf{A}-\textsf{D}} ~=~ -\,\frac{2}{(2N+1)^2}\,\sqrt{(N+\thalf)^2 - M_J^2}\,\,\,
{\cal B} \ ,
\label{appA6}
\end{equation}
where we define
\begin{equation}
{\cal B}~=~ \frac{1}{c_e(v,N)} \big(g_e + g_m(v,N)\big) \mu_B B \ .
\label{appA6a}
\end{equation}

This allows the $O(B)$ corrections to the results quoted in the text for $V_{\rm SME\,\pm}^e(R)$ to be
read off directly. For example, to calculate the $B$-dependence of the coefficients $\hat{c}_{NM_J}^\pm(B)$
in (\ref{b26a}),(\ref{b26b}) in the text, we apply this formula to the coefficient of 
$\VV_{220}^{e\,\NR}$ in $V_{\rm SME}^{e\,\VV}(R)$ where, taking out the common factor 
$\thalf \sqrt{\tfrac{5}{4\pi}}\, \tr_Y\la p_a\,p_b\ra\,\VV_{220}^{e\,\NR}$, we have from  Paper 2, Appendix B,
\begin{align}
\a ~&=~ \frac{1}{(2N+1)(2N+3)}\, \big[(N+\tfrac{1}{2}) (N+\tfrac{3}{2}) - 3 M_J^2\big] \ , \nonumber \\[4pt]
\d~&=~ \frac{1}{(2N-1)(2N+1)}\, \big[(N-\tfrac{1}{2}) (N+\tfrac{1}{2}) - 3 M_J^2\big] \ , \nonumber \\[4pt]
\b ~&=~ -6 \frac{1}{(2N-1)(2N+1)(2N+3)} \, \sqrt{(N+\tfrac{1}{2})^2 - M_J^2}\, M_J \ ,
\label{appA7}
\end{align}
and find
\begin{align}
\wh{c}_{NM_J}^{\,+}(B) ~&=~ \frac{1}{(2N+1)(2N+3)}\,\big[(N+\thalf)(N+\tfrac{3}{2}) - 3 M_J^2\big] \ ,  
\nonumber \\[8pt]
&~~~~~~~~~~~~~~~~~~~~~~~~~~~~~
+~24 \,\frac{\big[(N+\thalf)^2 - M_J^2\big]\,M_J}{(2N+1)^3 (2N-1) (2N+3)}\,\,\,
{\cal B}\ ,
\label{appA8}
\end{align}
and 
\begin{align}
\wh{c}_{NM_J}^{\,-}(B) ~&=~ \frac{1}{(2N-1)(2N+1)}\,\big[(N-\thalf)(N+\thalf) - 3 M_J^2\big] 
\nonumber \\[8pt]
&~~~~~~~~~~~~~~~~~~~~~~~~~~~~~ 
-~24 \,\frac{\big[(N+\thalf)^2 - M_J^2\big]\,M_J}{(2N+1)^3 (2N-1) (2N+3)}\,\,\,
{\cal B} \ .
\label{appA9}
\end{align}

\vskip0.3cm
In the large magnetic field limit, the eigenstates of the hyperfine-Zeeman Hamiltonian are $|v\,N\,M_J\,M_S\ra$
(or equivalently, $|v\,N\,M_N\,M_S\ra$ since $M_J=M_N + M_S$) so these are the appropriate states to use
for large fields. In this case the matrix elements $\la \,H_{HFS} + H_Z\,\ra$ are written for fixed $M_J$
in $2 \times 2$ matrix form with $M_S', M_S = \pm 1/2$ as,
\begin{equation}
\la v\,N\,M_J\,M_S'|\,H_{HFS} + H_Z\,|v\,N\,M_J\,M_S\ra ~=~ 
\begin{pmatrix}\textsf{A} ~&\textsf{B} \\ \textsf{B} ~&\,\textsf{D} \end{pmatrix} \ ,
\label{appA10}
\end{equation}
where now (see Paper 2, eq.(5.26),
\begin{align}
A ~&=~ \frac{1}{2}c_e(v,N)\,(M_J-\thalf) \,+\, \big( \thalf g_e - g_m(v,N)\big)\,(M_J-\thalf)\big)\,\mu_B B \ ,
\nonumber \\[4pt]
D ~&=~ -\frac{1}{2}c_e(v,N)\,(M_J+\thalf) \,-\, \big( \thalf g_e + g_m(v,N)\big)\,(M_J+\thalf)\big)\,\mu_B B \ ,
\nonumber \\[4pt]
B ~&=~ \frac{1}{2} c_e(v,N)\,\sqrt{(N+\thalf)^2 - M_J^2} \ .
\label{appA11}
\end{align}

The matrix elements $V_{\rm SME}^{e\,\VV}(R)$ for the spin-independent couplings are of course diagonal
in this basis,
so we just have $V_{\rm SME}^{e\,\VV}(R) \,=\, \begin{pmatrix} \a ~&0\\ 0 ~&\,\d \end{pmatrix} $
with the $O(p^2)$ terms,
\begin{align}
\a ~&=~ -\frac{1}{\sqrt{4\pi}}\,\bigg[ \tr\la p_a\,p_b\ra \VV_{200}^{e\,\NR} \,-\, \frac{\sqrt{5}}{2}\,\tr_Y\la p_a\,p_b\ra\,
c_{N\,M_J-\thalf}\, \VV_{220}^{e\VV} \,\bigg]\ ,  \nonumber\\[6pt]
\d ~&=~-\frac{1}{\sqrt{4\pi}}\,\bigg[ \tr\la p_a\,p_b\ra \VV_{200}^{e\,\NR} \,-\, \frac{\sqrt{5}}{2}\,\tr_Y\la p_a\,p_b\ra\,
c_{N\,M_J+\thalf}\, \VV_{220}^{e\VV}\,\bigg] \ .
\label{appA12}
\end{align}
In this case, we find the magnetic field corrections of $(c_e/\mu_B B)$ vanish, since $\b=0$, but we can use
the formula (\ref{appA5a}) to calculate the next order corrections. We find
\begin{equation}
c_{NM_J\mp\thalf} ~ \rightarrow ~ c_{NM_J\mp\thalf} ~\mp~ \frac{3}{2}\,\frac{[(N+\thalf)^2 - M_J^2]\,M_J }{(2N-1)(2N+3)}
\,\,\Big(\frac{c_e(v,N)}{(g_e+g_m(v,N))\mu_B B}\Big)^2  \ .
\label{appA13}
\end{equation}}

\vskip0.3cm
For the spin-dependent couplings, the matrix elements of $V_{\rm SME}^{e\,\TT}({\bs R})$ in the states
$|v\,N\,M_J\,M_S\ra$ take the form, with matrix rows/columns corresponding to $M_S=\pm \thalf$,
\begin{equation}
V_{\rm SME}^{e\,\TT}(R) ~=~ 
\begin{pmatrix}\a ~&\b \\ \b^* ~&\,\d \end{pmatrix} \ ,
\label{appA14}
\end{equation}
with 
\begin{align}
\a ~=~ &-\TT_{010}^{e\,\NR\,(0B)} ~-~\frac{1}{3}\,\tr\la p_a\,p_b\ra \,\,\big(\TT_{210}^{e\,\NR\,(0B)} + 
2\, \TT_{210}^{e\,\NR\,(1B)} \big)
\nonumber \\[6pt]
&+\,\frac{1}{3}\,\tr_Y\la p_a\,p_b\ra \, c_{N M_J-\thalf}\,\,
\big(\TT_{210}^{e\,\NR\,(0B)} - \TT_{210}^{e\,\NR\,(1B)} +\, \frac{3}{2}\,\sqrt{\frac{7}{6}}\, \TT_{230}^{e\,\NR\,(0B)} \big) \ ,
\label{appA15}
\\[10pt]
\d ~=~ &~~~\TT_{010}^{e\,\NR\,(0B)} ~+~\frac{1}{3}\,\tr\la p_a\,p_b\ra \,\,\big(\TT_{210}^{e\,\NR\,(0B)} + 
2\, \TT_{210}^{e\,\NR\,(1B)} \big)
\nonumber \\[6pt]
&-\,\frac{1}{3}\,\tr_Y\la p_a\,p_b\ra \, c_{N M_J+\thalf}\,\,
\big(\TT_{210}^{e\,\NR\,(0B)} - \TT_{210}^{e\,\NR\,(1B)} +\, \frac{3}{2}\,\sqrt{\frac{7}{6}}\, \TT_{230}^{e\,\NR\,(0B)} \big) \ ,
\label{appA16}\\[10pt]
\b ~=~&- \tr_Y \la p_a\,p_b\ra\,\,\,\frac{\sqrt{(N+\thalf)^2 - M_J^2}}{(2N-1)(2N+3)}\,\,M_J\,\,   
\nonumber \\[6pt]
&~~~~~~~~~\times\, 
\big( \TT_{210}^{e\,\NR\,(0B)} - \TT_{210}^{e\,\NR\,(1B)} -
\sqrt{\frac{7}{6}}\, \TT_{230}^{e\,\NR\,(0B)} -\sqrt{5}\,i\, \TT_{220}^{e\,\NR\,(1E)}\big) \ .
\label{appA17}
\end{align}

In this case, since $\b \neq 0$, there is a correction of $O(1/{\cal B})$ to the 
eigenvalues $V_{\rm SME\,\pm}^{e\,\TT}(R)$
and we can again  use the formula (\ref{appA5}), but this time for small $c_e/\mu_B B$ where we have from 
(\ref{appA11}),
\begin{equation}
\frac{\textsf{B}}{\textsf{A}-\textsf{D}} ~=~ \frac{1}{2}\, \sqrt{(N+\thalf)^2 - M_J^2}\,\,\,\frac{1}{{\cal B}} ~~ 
+~~ O(1/{\cal B})^2 \ .
\label{appA18}
\end{equation}
Together with the coefficients $\a,\,\d,\,\b$ above, we find $V_{\rm SME\,\pm}^{e\,\TT}(R)$ as given in the main
text, (\ref{c23}).

\section{Clebsch-Gordan relations}\label{appB}

Extending the set of identities given in Paper 2, Appendix B, to those required for the $O(p^4,P^4)$ SME contributions
considered here, we find 
\begin{align}
&\sum_{M_S}\, \Big(C_{N\,M_N,\,\thalf\,M_S}^{N+\thalf\,M_J}\Big)^2\,\,c_{N M_N}^{(4)}  ~~\equiv ~~ \hat{c}_{NM_J}^{(4)+}
\nonumber \\[4pt]
&=~\frac{3}{64}\,\, \frac{3(2N+5) (2N+3) (2N+1) (2N-1) - 40 (12N^2 + 24N -1) M_J^2 + 560 M_J^4}
{(2N+5)(2N+3)(2N-1)(2N-3)}  \ ,  \nonumber \\[15pt]
&\sum_{M_S}\, \Big(C_{N\,M_N,\,\thalf\,M_S}^{N-\thalf\,M_J}\Big)^2\,\,c_{N M_N}^{(4)}  ~~\equiv ~~ \hat{c}_{NM_J}^{(4)-}
\nonumber \\[4pt]
&=~ \frac{3}{64}\, \,\frac{3(2N+3)(2N+1)(2N-1)(2N-3) - 40 (12N^2 - 13) M_J^2 + 560 M_J^4}
{(2N+3)(2N+1)(2N-1)(2N-3)} \ ,   \nonumber \\[15pt]
&\sum_{M_S}\, C_{N\,M_N,\,\thalf\,M_S}^{N+\thalf\,M_J} \,\,C_{N\,M_N,\,\thalf\,M_S}^{N-\thalf\,M_J}
\,\,c_{N M_N}^{(4)}  \nonumber \\[4pt]
&=~ -\frac{15}{4} \,\,\frac{\big(12 N(N+1) -17 - 28M_J^2\big) \big[(N+\tfrac{1}{2})^2 - M_J^2\big] \,M_J}
{(2N+5)(2N+3)(2N+1)(2N-1)(2N-3)}  \ . \nonumber \\
\label{appB1}
\end{align}
where $c_{N M_N}^{(4)}$, defined as
\begin{equation}
C_{N M_N, 4 0}^{NM_N}\, C_{N 0,4 0}^{N 0} ~=~  c_{N M_N}^{(4)} \ ,
\label{appB2}
\end{equation}
is given explicitly by,
\begin{equation}
c_{NM_N}^{(4)} ~=~ \frac{3}{4} \,\,\frac{\big[3N(N+2)(N+1)(N-1) \,-\,5 (6N^2 + 6N -5) M_N^2 \,+\, 35 M_N^4 \big]}
 {\big[(2N+5)(2N+3)(2N-1)(2N-3)\big]}\ .
\label{appB3}
\end{equation}

\newpage

\section{Proton momentum expectation values}\label{appC}

In the main text, we quote the expectation values $\la v\,N|\,P^2\,|v\,N\ra$ and  $\la v\,N|\,P^4\,|v\,N\ra$
as expansions in $(v +\thalf)$ and $N(N+1)$, characterised by the small parameter
$\l \sim O(\sqrt{m_e/m_p})$.
Here we briefly illustrate a general method to find the expectation values of any operator in these
rovibrational states and apply it to the special case of $P^2$ and $P^4$. The method is essentially described
in Appendix B of Paper 2, which we use extensively though many of the results below are new. 

The eigenstates $|v\,N\ra$ of the proton Schr\"odinger equation are found by expanding about
the minimum $R_0$ of the inter-nucleon potential, treating the angular momentum-induced 
effective potential $V_N(R)$ as a perturbation. Recall that $V_M(R)$ is obtained by solving the
electron Schr\"odinger equation, while $V_N(R) = N(N+1)/2\mu R^2$. The expansion parameter
is $\l = 1/\mu\omega_0 R_0^2$. 

The unperturbed states are therefore those of a simple harmonic oscillator (SHO) with potential
$\thalf \mu\omega_0^2 \,x^2$, where $x = R-R_0$ and $\mu\omega_0^2 = V_M^{\prime\prime}(R_0)$.
The total perturbation, before including the SME, is $\d V = \d V_M + \d V_N$ where,
\begin{align}
\d V_M ~&=~ \frac{1}{6}\,V_M^{\prime\prime\prime}\, x^3 \,+\, \ldots \nonumber \\
\d V_N ~&=~ \l\,\omega_0\, N(N+1) \Big( - \frac{x}{R_0} \,+\, 
\frac{3}{2}\,\frac{x^2}{R_0^2} \,+\, \ldots\,\Big)  \ .
\label{appC1}
\end{align}
In what follows we present results up to 2nd order in perturbation theory, truncating the $\l$
expansion at $O(\l^2)$.  Higher order terms are calculated in Paper 1.

First consider the expectation value $\la v\,N|\,P^2\,|v\,N\ra$. 
In the nucleon Schrodinger equation the 3-dim momentum squared $P^2$ is represented as the
Lapalacian acting on the appropriate wave function,  with
\begin{align}
P^2 \,\rta\, - \nabla_{\bs R}^2 ~&=~ - \frac{d^2}{dR^2} \,+\, N(N+1)\,\frac{1}{R_0^2} \nonumber \\[4pt]
&=~~ k^2 \,+\, N(N+1)\,\frac{1}{R_0^2}\,\Big( 1 \,-\, 2\frac{x}{R_0} \,+\, 
3 \frac{x^2}{R_0^2} \,+\,\ldots \,\Big) \ ,
\label{appC2}
\end{align}
where we denote the 1-dim SHO momentum by $k = -i d/dx$.
We therefore need to find the expectation values of the SHO operators $k^2$, $\,x$ and $x^2$ 
in the states $|v\,N\ra$ up to 2nd order in the perturbations $\d V$. Taking intermediate results
from Paper 1, we find:
\begin{align}
\la\,k^2\,\ra ~&=~ \m \omega_0 \, (v+\thalf) \,\Big( 1\,+\, \frac{3}{2} \,N(N+1)\,\l^2\Big) 
\label{appC3} \\[4pt]
\frac{1}{R_0^3}\,\la x\,\ra ~&=\, -\frac{1}{2} \,\m\omega_0\,(v+\thalf)\, 
\Big(\frac{R_0 V_M^{\prime\prime\prime}}{V_M^{\prime\prime}}\Big)\,\l^2 \ ,
\label{appC4} \\[4pt]
\frac{1}{R_0^4}\,\la\,x^2\,\ra ~&=~ \m\omega_0\,(v+\thalf)\,\l^2 \ .
\label{appC5} 
\end{align}
Collecting terms, recalling $\m = \thalf m_p$, and truncating at $O(\l^2)$ gives\footnote{The coefficient 
of the $V_M^{\prime\prime\prime}$
term in (\ref{appC6}) differs from what would be expected from a simple scaling argument used in Paper 1,
which would give a factor $3/2$ rather than 1. We have been unable to resolve this discrepancy, though
one possibility is that there could be a further contribution at 3rd order in perturbation theoryin addition
to that calculated above. 
With this caveat, we have therefore chosen to quote the original `scaling' coefficient in (\ref{d16}) in 
the main text rather than (\ref{appC6}).   We have also omitted $O(\l)$ terms proportional to $(v+\thalf)^2$
contributing to the $x_{\rm SME}^n$ coefficient; these depend on the third and fourth derivatives of $V_M$
and are quoted in Paper 1.}
\begin{align}
\la v\,N|\,P^2\,|v\,N\ra ~=~ \half m_p \,\omega_0\,\bigg[(v+\thalf) \,&+\, N(N+1)\,\l  \nonumber \\[2pt]
&\,+\, \Big(\frac{9}{2} \,+\,\frac{R_0 V_M^{\prime\prime\prime}}{V_M^{\prime\prime}}\Big)
 \,(v+\thalf)\,N(N+1) \,\l^2 ~+\ldots \bigg] \ .
\label{appC6}
\end{align}

\vskip0.3cm
The analysis of $\la v\,N|\,P^4\,|v\,N\ra$ follows along the same lines, though is inevitably more complicated.
Here, in the nucleon Schr\"odinger equation,
\begin{align}
P^4 \,\rta\,  \nabla_{\bs R}^2\,\nabla_{\bs R}^2 ~&=~  \frac{d^4}{dR^4} \,-\,
N(N+1)\,\frac{1}{R^2}\,\Big( 2\, \frac{d^2}{dR^2} \,-\, \frac{4}{R}\,\frac{d}{dR} \,+\, \frac{6}{R^2}\,\Big) 
\,+\, \big(N(N+1)\big)^2 \,\frac{1}{R^2} \nonumber \\[6pt]
&=~~~k^4 \,+\, N(N+1)\,\frac{1}{R_0^2}\,\Big( 2\,k^2\big(1 + 3\frac{x^2}{R_0^2}\big)
\,+\, 4 i k \frac{1}{R_0}\,\big(1 - 3\frac{x}{R_0}\big) \,-\,\frac{6}{R_0^2} \Big) \nonumber \\[6pt]
&~~~~~~~~~~~~~~~~~~~~+\, 
\big(N(N+1)\big)^2 \frac{1}{R_0^4} ~+\,\ldots 
\label{appC7}
\end{align}
keeping only terms which give relevant contributions at the order considered here.
Again working up to 2nd order in perturbation theory making extensive use of results from Paper 1,
and truncating at $O(\l^2)$, we find
\begin{align}
\la\,k^4\,\ra ~&=~ \frac{3}{2}\,(\m \omega_0)^2 \, \big[(v+\thalf)^2 + \tfrac{1}{4}\big] \,
\Big(1 \,+\,\frac{3}{2}\,N(N+1)\,\l^2 \Big) \ ,
\label{appC8} \\
\frac{1}{R_0^4}\,\la\,k^2\,x^2\,\ra ~&=~ \frac{1}{2}\,(\m \omega_0)^2\,
\big[(v+\thalf)^2 \,-\,\tfrac{3}{4}\big]\,\l^2 \ ,
\label{appC9} \\
\frac{1}{R_0^3}\,\la\,i k\,\ra ~&=~ O(\l^3) \ ,
\label{appC10} \\
\frac{1}{R_0^4}\,\la\,i k\,x\,\ra ~&=~\frac{1}{2}\, (\m \omega_0)^2\,\l^2 \ ,
\label{appC11}
\end{align}
in addition to those quoted above. Collecting all these terms, we obtain
\begin{align}
\la v\,N|\,P^4\,|v\,N\ra ~=~ (\m \omega_0)^2 \,\,\bigg[ &\frac{3}{2}\,\big[(v+\thalf)^2 + \tfrac{1}{4}\big]
\,\Big(1\,+\,\frac{3}{2}\,N(N+1)\,\l^2\Big)   \nonumber \\[2pt]
&+\, 2\,(v+\thalf)\,N(N+1)\,\l\, \Big(1 \,+\,\frac{3}{2}\,N(N+1)\,\l \Big)\nonumber \\[4pt]
&+\, 3\,\big[(v+\thalf)^2 \,-\,\tfrac{3}{4}\big]\,N(N+1)\, \l^2 \nonumber \\[6pt]
&-\, N(N+1)\,\big( 12 \,-\, N(N+1)\big)\,\l^2 ~+\,\ldots \ .
\label{appC12}
\end{align}
Omitting the terms of $O\big((N(N+1)\big)^2$ and $O\big((v+\thalf)^2\,N(N+1)\big)$, 
this is the result quoted in the main text, (\ref{d23}).
Notice the unusual looking coefficient of $N(N+1)$ which appears there. This arises because
of contributions from the $O(1)$ fractions which automatically accompany the terms of $O(v+\thalf)^2$.

\end{document}